\documentclass[letterpaper,twocolumn,10pt]{article}
\usepackage{usenix2019_v3}

\usepackage{tikz}
\usepackage{amsmath}
\usepackage{amsfonts}
\usepackage{booktabs}
\usepackage{titling}

\usepackage{enumitem}
\setlist[itemize]{leftmargin=0cm,itemindent=.5cm,labelwidth=0.6\itemindent,labelsep=0cm,align=left, itemsep=.0cm, topsep=0cm}
\setlist[enumerate]{leftmargin=0cm,itemindent=.3cm,labelwidth=0.6\itemindent,labelsep=0cm,align=left, itemsep=.0cm, topsep=0cm}

\usepackage{filecontents}
\usepackage[linesnumbered, ruled, vlined]{algorithm2e}
\usepackage{subcaption}
\usepackage{enumitem}
\usepackage{xspace}
\usepackage{amsthm}

\newcommand{\parab}[1]{\noindent\textbf{#1}}

\newif\ifprintcomments
\printcommentstrue

\newcommand{\epname}[0]{FineEP\xspace}
\newcommand{\sysname}[0]{FineMoE\xspace}

\begin{document}

\date{}

\title{\Large \bf \sysname: Fine-grained Load Balancing for Mixture-of-Experts with\\Token Scheduling}

\author{
Chenqi Zhao\\
Peking University
\and
Wenfei Wu\\
Peking University
\and
Linhai Song\\
Institute of Computing Technology, Chinese\\ Academy of Sciences
\and
Yuchen Xu\\
Peking University
\and
Yitao Yuan\\
Peking University
}

\maketitle

\begin{abstract}
Mixture-of-Experts (MoE) has emerged as a promising approach to scale up deep learning models due to its significant reduction in computational resources.
However, the dynamic nature of MoE leads to load imbalance among experts, severely impacting training efficiency.
While previous research has attempted to address the load balancing challenge, existing solutions either compromise model accuracy or introduce additional system overhead.
As a result, they fail to achieve efficient and fine-grained load balancing, which is crucial to optimizing training efficiency.

We propose \epname, a novel parallelization strategy to achieve fine-grained load balancing in MoE systems.
\epname is capable of achieving complete load balancing in every micro-batch through efficient token scheduling across GPUs.
Furthermore, we propose \sysname, an efficient distributed MoE training system with \epname's load balancing capabilities.
Our experimental results demonstrate that \sysname improves the end-to-end training throughput by up to 47.6\% compared with the state-of-the-art system, and almost consistently achieves complete load balance among GPUs.
\end{abstract}

\section{Introduction}
\label{sec:introduction}
In recent years, transformer-based large models have fueled significant advancements across various domains, including natural language processing~\cite{Transformer, GPT-2, PaLM}, computer vision~\cite{V-MoE, Swin_Transformer}, and multimodality~\cite{ViLBERT, Cao_2023_ICCV}.
These achievements underscore that scaling up model size is a straightforward yet fundamental approach to enhancing model performance. 
Nevertheless, deploying large models necessitates substantial computational resources.
To tackle this, Mixture-of-Experts (MoE) proposes a sparsely-activated architecture~\cite{shazeer2017outrageouslylargeneuralnetworks}, enabling the scaling up of model capacity with only a sub-linear increase in computational cost.
Nowadays, MoE has gained widespread adoption in state-of-the-art models, such as GPT-5~\cite{GPT-5}, Llama 4~\cite{Llama-4}, Gemini 3~\cite{Gemeni-3}, Grok-4~\cite{Grok-4}, Claude 4~\cite{Claude4}, Qwen3~\cite{Qwen3}, and DeepSeek-V3~\cite{DeepSeek-V3}.

However, the sparse nature of MoE can indeed introduce inefficiencies during training.
MoE divides a model layer into multiple \emph{experts} and routes each input token to its \emph{top-K} most suitable experts at runtime.
This dynamic routing can result in significant load imbalances among experts~\cite{FasterMoE, GShard}.
Furthermore, recent distributed training frameworks commonly employ \emph{expert parallelism} (EP) to distribute experts across multiple GPUs~\cite{Switch_Transformer}.
While EP reduces memory consumption per GPU, it introduces a \emph{straggler} problem:
All GPUs must wait for the most heavily loaded GPU in the EP group to complete computation before proceeding, resulting in wasted computational resources and reduced system throughput.
Therefore, the load imbalance issue poses a significant challenge in efficiently training MoE models.

To address the load imbalance issue of MoE, many works propose \emph{algorithmic solutions} by modifying the token-to-expert routing algorithm.
These approaches incorporate load-balancing loss~\cite{Switch_Transformer, GLaM, auxiliary-loss-free}, route tokens to less suitable experts~\cite{LocMoE, expert_choice}, or simply drop excess tokens to alleviate the burden on heavily loaded experts~\cite{Switch_Transformer, GShard, LocMoE}.
However, these modifications to model logic can degrade model accuracy and potentially impact model convergence~\cite{guo2025advancingexpertspecializationbetter, OpenMoE, Prophet}.
As a result, they introduce a trade-off between model accuracy and system efficiency~\cite{Pangu, qiu2025demonsdetailimplementingload}. 

Other works propose \emph{systematic solutions} for load balancing in MoE without compromising the model accuracy.
A series of works balance GPU loads through \emph{expert scheduling}, adjusting the expert-to-GPU mapping~\cite{SmartMoE, FlexMoE, Hecate, Pro-Prophet, Prophet, SwiftMoE}.
For instance, FlexMoE~\cite{FlexMoE} adaptively replicates experts based on their popularity, distributing loads of more popular experts to more GPUs.
However, these solutions cannot effectively achieve ideal load balance, which should satisfy two fundamental requirements:

\textbf{R1: Maximizing system efficiency requires fine-grained load balance among all GPUs.}
Giant model sizes demand numerous GPUs for training~\cite{Pangu, DeepSeek-V3}, making even minor load imbalances costly in terms of wasted GPU time.
The significant expense necessitates a fine granularity for load balancing, hopefully equalizing all GPU loads.
However, existing expert scheduling solutions typically treat each expert replica as a scheduling unit for communication efficiency~\cite{SmartMoE, FlexMoE, Hecate, Prophet}.
Since scheduling at the expert granularity results in a limited, discrete scheduling space, they fail to achieve optimal load balance across a vast, near-continuous space of possible load distributions.

\textbf{R2: Dynamic expert load distributions require micro-batch-level adaptation at runtime.}
Data heterogeneity across micro-batches causes severe variations in expert loads, necessitating dynamic load balancing to optimize GPU utilization consistently.
Existing expert scheduling solutions involve migrating expert parameters to align with the changing load distributions~\cite{FlexMoE, Prophet, SmartMoE}.
However, the substantial size of these parameters presents a significant challenge to making frequent adjustments seamlessly.

{To meet these requirements, }
we adopt a novel systematic approach by \textbf{\textit{scheduling input tokens}} instead of experts, which has two advantages:
(\textbf{R1}) First, token scheduling allows a precise control over per-GPU workloads (relative to token counts), enabling fine-grained load balancing.
(\textbf{R2}) Second, token scheduling leverages existing all-to-all communication in EP, incurring minimal communication overhead.
This allows us to achieve dynamic load balancing within each micro-batch without compromising the entire system efficiency.
Furthermore, we can combine fine-grained token scheduling and coarse-grained expert scheduling, improving overall load balancing capability.

However, performing token scheduling in existing EP frameworks faces two challenges.
(\textbf{C1}) Current frameworks dispatch tokens within an EP group, which contains exactly one replica of each expert~\cite{Megatron-LM, DeepSpeed}.
As a result, the GPU loads are fixed by expert loads, which are determined by the routing algorithm.
To tackle this, we schedule tokens among expert replicas across \emph{multiple EP groups}, balancing GPU loads while keeping expert loads.
(\textbf{C2}) Identical expert placement across EP groups severely limits the scheduling space (see \figurename~\ref{fig:overview-2}).
Since replicas of a popular expert have identical EP ranks in all EP groups, na\"ive token scheduling cannot migrate their loads to other EP ranks, resulting in only local load balance.
Although some studies adopt non-identical expert placement~\cite{FlexMoE}, the relationship between expert placement and token scheduling remains uninvestigated.
To tackle this, we theoretically analyze the influence of expert placement on token scheduling with \emph{graph theory} (\S\ref{ssec:placement_analysis}).
Based on the analysis, we tailor the \emph{expert placement} to expand the scheduling space and achieve global load balance.

We propose \epname, a novel EP strategy that leverages token scheduling for fine-grained load balancing across GPUs.
\epname optimizes load balancing from two perspectives:
(1) From a short-term perspective, \epname schedules tokens within every micro-batch using \emph{linear programming} to minimize the maximum GPU loads (\S\ref{ssec:lpp}). 
To optimize system efficiency, \epname reduces communication overhead through \emph{locality-aware routing} (\S\ref{ssec:routing}) and minimizes scheduling latency through \emph{overlapping} (\S\ref{ssec:when}).
(2) From a long-term perspective, \epname tailors \emph{expert placement} to optimize its token scheduling performance and mitigate coarse-grained imbalances.
Based on our theoretical analysis, \epname incorporates multiple expert placement strategies (\S\ref{ssec:symmetric}-\ref{ssec:asymmetric}) and an \emph{adaptive replacement} mechanism (\S\ref{ssec:adaptive}).
These optimizations enable \epname to maintain complete load balance under dynamic and highly skewed workloads.

Based on \epname, we design \sysname, an efficient distributed MoE training system.
We implement \sysname upon Megatron-LM~\cite{Megatron-LM} and evaluate its load balancing performance.
Experimental results show that \sysname achieves up to 47.6\% improvement on end-to-end throughput compared with Megatron-LM and maintains complete load balance even under highly imbalanced workloads.

Briefly, we make the following contributions:

\begin{itemize}
    \item We propose \epname to achieve dynamic and fine-grained load balancing among GPUs with token scheduling (\S\ref{ssec:overview}).

    \item We formulate the load balancing problem into a linear programming problem and solve it efficiently (\S\ref{sec:schedule}).

    \item We propose several expert placement strategies and an adaptive replacement mechanism to optimize \epname's load balancing capacity (\S\ref{sec:placement}).

    \item We implement \sysname, an efficient distributed MoE training system based on \epname.
    We conduct comprehensive experiments to demonstrate that \sysname exhibits superior system efficiency and load balancing capability (\S\ref{sec:evaluation}).
\end{itemize}

\section{Background}
\subsection{Transformer and MoE}
\begin{figure}
    \centering
    \begin{subfigure}{\linewidth}
    \centering
    \includegraphics[width=0.85\linewidth, trim=0 285 0 0, clip]{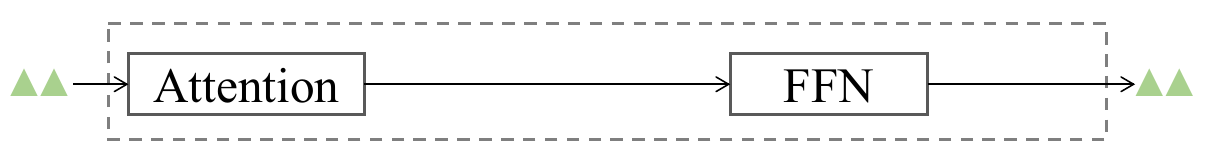}
    \caption{A dense transformer layer.}
    \label{fig:EP-1}
    \end{subfigure}
    \begin{subfigure}{\linewidth}
    \centering
    \includegraphics[width=0.85\linewidth, trim=0 180 0 90, clip]{figs/EP.pdf}
    \caption{An MoE transformer layer.}
    \label{fig:EP-2}
    \end{subfigure}
    \begin{subfigure}{\linewidth}
    \centering
    \includegraphics[width=0.9\linewidth, trim=0 5 0 180, clip]{figs/EP.pdf}
    \caption{An MoE transformer layer with expert parallelism.}
    \label{fig:EP-3}
    \end{subfigure}
    \caption{An example of transformer, MoE, and expert parallelism.}
    \label{fig:EP}
\end{figure}
Transformer is the state-of-the-art architecture for deep learning models~\cite{Transformer}.
A transformer model consists of multiple sequential layers, each comprising an attention layer and a feed-forward network (FFN). \figurename~\ref{fig:EP-1} illustrates the architecture of a standard (dense) transformer layer.

\figurename~\ref{fig:EP-2} illustrates an MoE transformer layer~\cite{Switch_Transformer}.
MoE replaces the FFN with multiple \emph{expert} FFNs, coordinated by a \emph{gate} network.
After the attention computation, the gate network assigns each token to its \emph{top-K} most suitable experts.
Then, these selected experts process the token and aggregate their results to produce the final result.

\subsection{Expert Parallelism and Expert Data Parallelism}
The increasing scale of deep learning models has necessitated the development of various parallelization strategies for distributed training across multiple computational devices.
We introduce some of them as follows.

\textit{Data parallelism (DP)} is a fundamental parallelization strategy that partitions input data across devices to enhance training throughput.
In the forward pass, each DP rank (e.g., a device) maintains a complete copy of the model parameters while processing distinct \emph{micro-batches} of data independently.
In the backward pass, all devices in a DP group aggregate their gradients for parameter synchronization.
Recent advances, such as ZeRO~\cite{ZeRO}, have further optimized DP's memory efficiency by eliminating memory redundancies across devices.

\textit{Tensor parallelism (TP)} addresses per-device memory constraints by partitioning model parameter tensors across devices. 
Each TP rank processes complete inputs using its allocated slice of model parameters, with results aggregated within the TP group to complete the computation.

\textit{Pipeline parallelism (PP)} addresses memory constraints by partitioning model layers~\cite{PipeDream, GPipe}.
While PP introduces lower communication overhead compared with TP, it wastes some computational resources due to pipeline bubbles~\cite{GPipe, Zero_bubble}.

\textbf{\textit{Expert parallelism (EP)}}
is a widely employed technique in distributed MoE systems~\cite{Switch_Transformer}.
EP is a specialized hybrid parallelization strategy combining elements of DP and TP.
By exploiting the inherent sparsity of MoE, EP benefits from both DP's training efficiency and TP's memory efficiency.
As shown in \figurename~\ref{fig:EP-3}, EP partitions the FFN layer by distributing distinct experts across multiple devices.
At runtime, the attention layer and the gate network utilize conventional DP to process different tokens on separate devices.
Next, devices within the EP group perform an all-to-all communication operation to dispatch tokens towards their designated expert devices.
Then, each device processes its received tokens with its local experts.
Finally, the devices perform another all-to-all operation to return tokens to their original devices.

\textbf{\textit{Expert data parallelism (EDP)}} is a specialized form of DP emerging from EP.
In configurations where the DP degree (number of DP ranks) exceeds the EP degree, systems must employ EDP across multiple EP groups within a DP group.
Each EDP group consists of devices sharing the same EP rank across different EP groups.
Devices within an EDP group maintain replicated instances of identical experts while processing distinct tokens from different EP groups.
The synchronization of parameters and gradients among these expert replicas follows conventional DP mechanisms, ensuring consistency across the distributed system. 

\subsection{Load Imbalances in MoE}

\begin{figure}[t]
    \centering
    \includegraphics[width=\linewidth, trim=10 30 10 10]{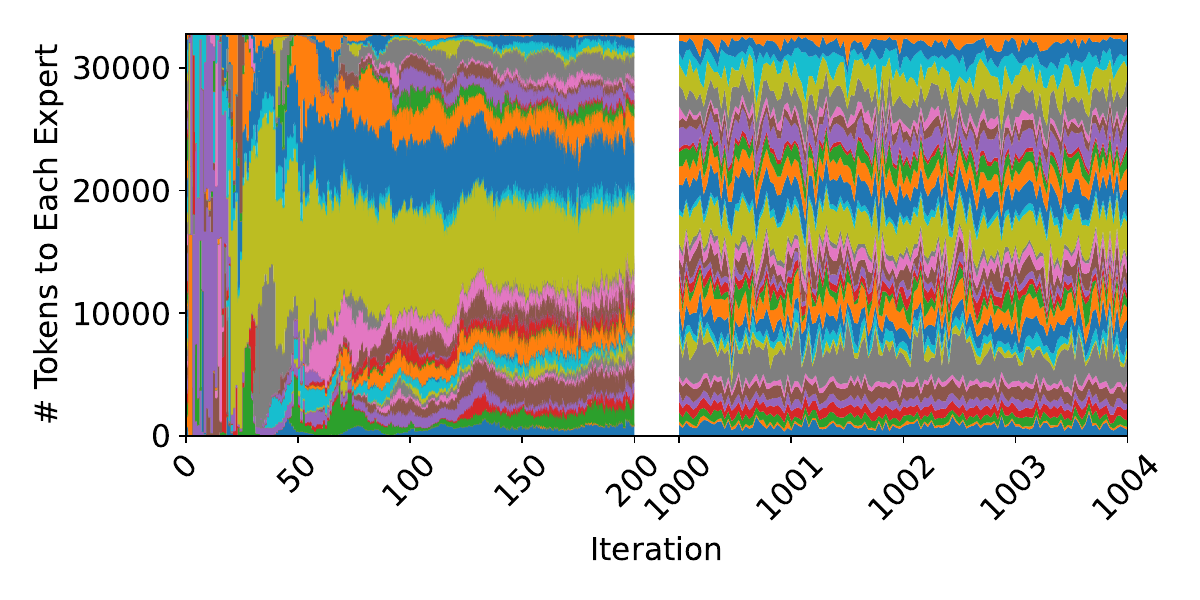}
    \caption{Expert load distribution of GPT 32$\times$1.3B layer 20 in some training iterations.}
    \label{fig:dist}
\end{figure}
The dynamic token-to-expert assignment in MoE systems inherently results in significant load imbalances across experts, presenting a critical performance bottleneck for distributed training.
We conduct an experiment to trace the expert load distribution in a training process. 
As illustrated in the left part of \figurename~\ref{fig:dist}, \textbf{the expert load distribution exhibits substantial variability and skewness}, particularly during the initial training iterations.

The imbalance issue severely impacts resource utilization and training throughput in MoE systems with EP.
In current MoE implementations, the FFN computation time of a GPU is approximately proportional to the total number of tokens assigned to the experts on this GPU~\cite{Megablocks}.
Since GPUs within an EP group must synchronize through all-to-all communication both before and after FFN computation, all GPUs must wait until the slowest GPU completes computation before proceeding.
Consequently, the system's overall throughput is bottlenecked by the most heavily loaded device, namely the \emph{straggler}. 
This strict synchronization requirement emphasizes the necessity of effective load balancing strategies to optimize MoE training efficiency.

Moreover, the straggler effect continuously degrades training efficiency in \emph{every micro-batch}.
As shown in the right part of \figurename~\ref{fig:dist}, \textbf{expert load distribution fluctuates significantly between consecutive micro-batches}.
This dynamic nature of load imbalances necessitates fine-grained load balancing at the micro-batch level to maintain optimal training efficiency.

\section{Challenges in Token Scheduling}
\label{sec:overview}
\label{ssec:analysis}

\begin{figure}[t]
    \centering
    \begin{subfigure}{\linewidth}
    \includegraphics[width=\linewidth, trim=0 288 0 0, clip]{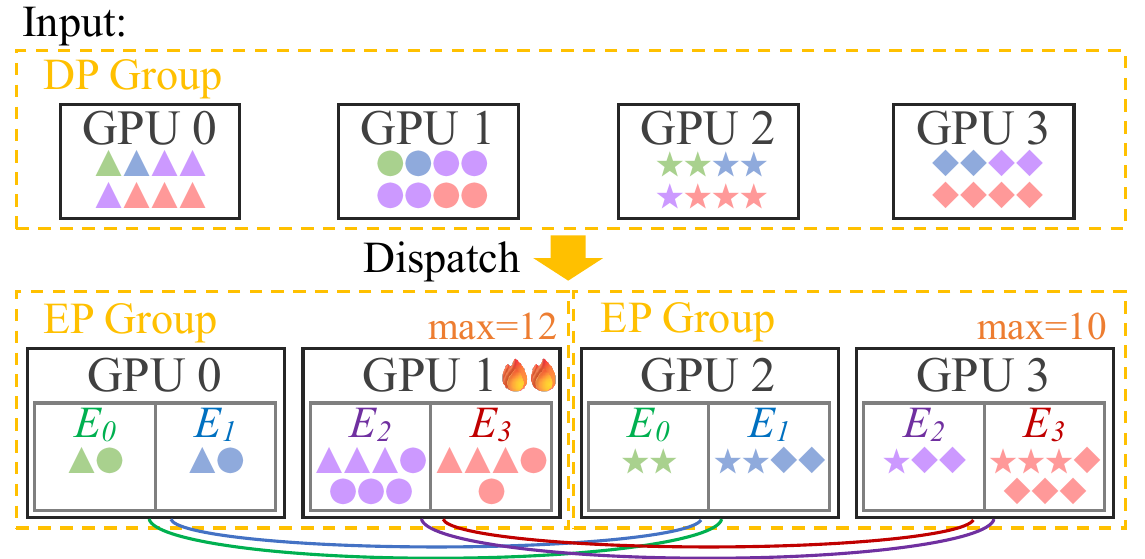}
    \caption{Vanilla EP. The DP degree is 4, and the EP degree is 2.}
    \label{fig:overview-1}
    \end{subfigure}
    \begin{subfigure}{\linewidth}
    \includegraphics[width=\linewidth, trim=0 152 0 270, clip]{figs/overview.pdf}
    \caption{Merging two EP groups. Afterward, GPU loads within EDP groups $\{0,2\}$ and $\{1,3\}$ are equal.}
    \label{fig:overview-2}
    \end{subfigure}
    \begin{subfigure}{\linewidth}
    \includegraphics[width=\linewidth, trim=0 1 0 410, clip]{figs/overview.pdf}
    \caption{\epname, shuffling expert placement and scheduling replica loads.
    Afterward, all GPU loads are equal.}
    \label{fig:overview-3}
    \end{subfigure}
    \caption{Converting EP to \epname. The shape and color of a symbol indicate the source GPU and the assigned expert of a token. The bottom curves indicate EDP groups.}
    \label{fig:overview}
\end{figure}


We aim to achieve fine-grained load balancing for MoE systems through \emph{token scheduling} within every micro-batch.
However, this approach poses two fundamental challenges:

\parab{Challenge 1:} \emph{Vanilla EP restricts token dispatching within each EP group, providing no space for scheduling.}

To perform token scheduling for load balancing, we must first determine the \emph{scheduling space}.
Specifically, when a token is assigned to an expert, we must be able to compute it on one of \emph{multiple} GPUs.

Unfortunately, such scheduling space is unattainable in the vanilla EP paradigm.
As shown in \figurename~\ref{fig:overview-1}, vanilla EP restricts token dispatching within the scope of an EP group, where each EP group contains exactly one replica of each expert's parameters.
When a token is assigned to an expert, it must be computed on the GPU hosting that expert's replica in its EP group.
Consequently, the token-to-GPU mapping is fixed by the token-to-expert mapping, eliminating any space for computation scheduling.

\parab{Solution 1:} \emph{Our key observation is that we can find scheduling space through \textbf{expert data parallelism}.}
We call the set of GPUs that host replicas of expert $e$ as the EDP group of expert $e$.
Since all replicas of an expert maintain identical parameters, a token can be computed equivalently with any replica in its designated expert's EDP group.
This observation inspires us to merge multiple EP groups and schedule tokens across their experts' EDP groups, as shown in \figurename~\ref{fig:overview-2}.

\parab{Challenge 2:} \emph{Identical expert placement across EP groups results in constrained scheduling space.}

While this approach improves token distribution, it falls short of optimal load balancing.
The fundamental limitation lies in the restricted scheduling space---load balancing occurs only within individual EDP groups.
In the current EP paradigm, each EP group maintains identical expert placement.
Consequently, the EDP groups of different experts are either completely disjoint or identical 
(e.g., in \figurename~\ref{fig:overview-2}, the EDP groups of expert 0,1 are both GPU \{0,2\}, while the EDP groups of expert 2,3 are both GPU \{1,3\}). 
This constraint indicates that we can at most equalize workloads within each EDP group, while significant load imbalances may persist across different EDP groups.

\parab{Solution 2:} \emph{To expand the scheduling space, we \textbf{shuffle the expert placement} within the merged EP groups}.
This operation creates intersecting EDP groups across different experts, substantially enlarging the scheduling space.
As illustrated in \figurename~\ref{fig:overview-3}, this approach can potentially achieve complete load balance across all GPUs.

\section{\sysname Overview}
\label{ssec:overview}
\begin{figure}
    \centering
    \includegraphics[width=\linewidth]{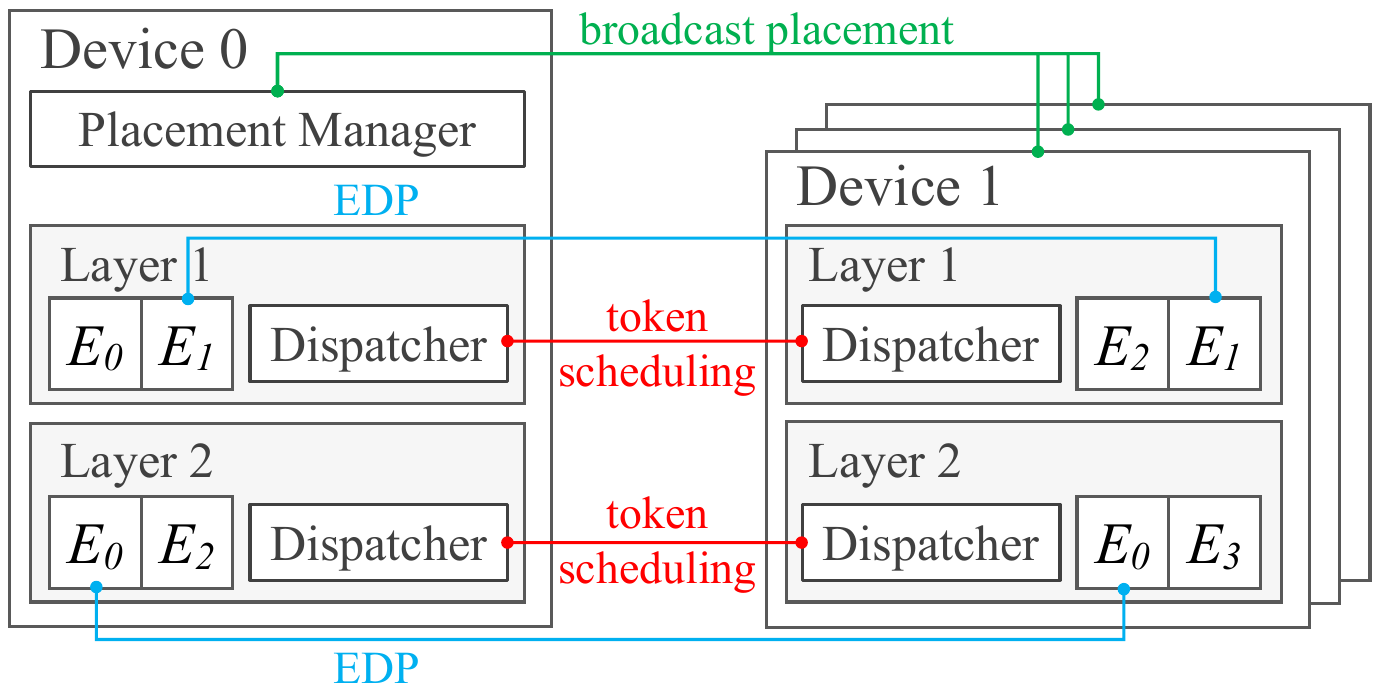}
    \caption{\sysname architecture.}
    \label{fig:system}
\end{figure}
We introduce a novel expert parallelism strategy, namely \epname.
\epname optimizes load balancing across GPUs within \emph{every micro-batch} through \emph{token scheduling} across \emph{EDP groups}.

Based on \epname, we propose \sysname, an efficient distributed MoE training system.
\figurename~\ref{fig:system} illustrates the architecture of \sysname.
Following PP, each device hosts multiple model layers.
Each MoE layer consists of a \emph{token dispatcher} and multiple local expert replicas.
Expert replica placement is coordinated by a global \emph{placement manager} residing on device 0.
We briefly illustrate the workflow of \sysname as follows.

\parab{Prerequisites.} 
\epname obtains scheduling space from intersecting EDP groups, which require two conditions: the DP degree must exceed the EP degree, and each GPU must host multiple expert replicas.
We find that these requirements are satisfied in many open-sourced MoE models~\cite{Pangu, DeepSeek-V3}.
For example, DeepSeek-V3's pre-training configuration employs 64-way EP and 128-way DP, with each GPU hosting 8 experts~\cite{DeepSeek-V3}. 

\parab{Configuration.} \epname introduces an integer parameter $d$, where $1<d\le \frac {\text{DP\_degree}} {\text{EP\_degree}}$.
\epname merges every $d$ EP groups into a \epname group.
In the following sections, we consider the simplest case where $d=\frac {\text{DP\_degree}} {\text{EP\_degree}}$, i.e., the entire DP group forms a single \epname group.

\parab{Initialization.} 
During model initialization, the placement manager generates expert placements within all \epname groups and broadcasts the placements to all devices. 
The placement strategies are detailed in \S\ref{sec:placement}.

\parab{Runtime.} During model training, \epname balances GPU workloads by scheduling tokens within the \epname group.
Specifically, after the gate networks assign tokens to specific experts, 
the dispatchers route tokens to particular expert replicas within their respective EDP groups.
The scheduling algorithm is detailed in \S\ref{sec:schedule}.

After scheduling, the dispatchers initiate an all-to-all (dispatch) operation within the \epname group, sending tokens to their routed expert replicas.
Then, each GPU executes FFN computation to process received tokens using its local expert replicas.
Afterward, a second all-to-all (combine) operation returns processed tokens to their source GPUs.
These operations follow the same patterns as vanilla EP, but occur in a communication group $d$ times larger.

\parab{An example.}
Surprisingly, \epname can almost always achieve complete load balance among GPUs, unless the expert load distribution is extremely skewed (as shown in \S\ref{ssec:vary_skew}).
We can intuitively learn how \epname achieves such good performance from the example in \figurename~\ref{fig:overview-3}.
Since expert 3 is heavily loaded, it places 8 tokens on GPU 3, preventing expert 0 from placing any token on the same GPU.
Fortunately, expert 0 can place all its 4 tokens on GPU 0.
Similarly, expert 2 also ``shifts'' some load from GPU 2 to GPU 1, while expert 1 shifts some load from GPU 1 to GPU 0.
Consequently, the computational burden of heavily loaded experts is effectively distributed across the entire \epname group.

\section{Token Scheduling}
\label{sec:schedule}
Token scheduling is the core of \epname for fine-grained load balancing.
The scheduling algorithm runs in two steps: The first step is to distribute expert loads among their replicas, balancing GPU loads (\S\ref{ssec:lpp}). The second step is to route tokens to specific expert replicas, enforcing the calculated replica loads (\S\ref{ssec:routing}).
Then, we discuss where and when to launch the scheduling in \S\ref{ssec:where} and \S\ref{ssec:when}.
Finally, we discuss some additional scheduling optimizations in Appendix~A.

\subsection{Determining Replica Loads}\label{ssec:lpp}

We aim to balance GPU loads through optimal distribution of expert loads among their replicas.
We formulate the load balancing problem as an optimization problem.

\begin{table}[tb]
\caption{List of Symbols and Notations.}
\begin{center}
\begin{tabular}{|l|l|}
\hline
\textbf{Symbol}&\textbf{Description} \\
\hline
$E$ & set of all experts \\
\hline
$G_{\epname}$ & set of GPUs in the concerned \epname group \\
\hline
$d$ & a parameter $=\frac{|G_{\epname}|}{\text{EP\_degree}}$ \\
\hline
$G_{EDP}^e$ & EDP group of expert $e$ \\
\hline
$load_e$ & total load of expert $e$ \\
\hline
$x_e^g$ & replica load of expert $e$ on GPU $g$ \\
\hline
$input_e^g$ & input load of expert $e$ from GPU $g$ \\
\hline
$m$ & optimal objective value of LPP~\ref{eq:optimization} \\
\hline

\end{tabular}
\label{tab:notation}
\end{center}
\end{table}

\begin{itemize}
    \item \emph{Notations}.
    Let $E$ represent the set of all experts, $G_{\epname}$ represent the set of GPUs in the concerned \epname group, and $G_{EDP}^e$ represent the set of GPUs in the EDP group of expert $e$.
    Since we currently focus on a single \epname group, we assume that $G_{EDP}^e\subseteq G_{\epname}, \forall e\in E$.
    
    \item \emph{Variables}.
    The variables are $\{x_e^g: e\in E, g\in G_{EDP}^e\}$, where $x_e^g$ is the replica load of expert $e$ on GPU $g$.

    \item \emph{Constraints}.
    Let $load_e (e\in E)$ denote the total load (number of tokens) of expert $e$ in the concerned \epname group.
    A valid distribution must ensure that each expert distributes its total load across its replicas.

    \item \emph{Objective}.
    Due to synchronization requirements for all-to-all communication before and after expert computation, the GPU with the highest load becomes the performance bottleneck.
    Therefore, we should minimize the maximum GPU load across the \epname group. 
\end{itemize}

We formulate the optimization problem as follows.
\begin{equation}
\begin{aligned}
    \text{minimize} & \max\limits_{g\in G_{\epname}}\left\{\sum\limits_{e\in E: g\in G_{EDP}^e}x_e^g\right\}, \\
    \text{subject to} & \sum\limits_{g\in G_{EDP}^e}x_e^g=load_e, \quad \forall e\in E, \\
    & x_e^g \ge 0, \quad \forall e\in E, g\in G_{EDP}^e.
    \label{eq:optimization}
\end{aligned}
\end{equation}

Problem \ref{eq:optimization} is a \emph{linear programming problem} (LPP), which can be efficiently solved in polynomial time relative to the number of GPUs and experts.
The number of variables is $O(|E|d)$, and the number of constraints is $O(|E|+|G_{\epname}|)$.
Given its modest scale, we solve this LPP using a single CPU thread with the HiGHs solver~\cite{HiGHS}.
GPU acceleration or multi-threading would not yield significant performance benefits for this scale of optimization.

Notably, across different micro-batches, while the constraint matrix (determined by expert placement $G_{EDP}^e$) remains the same, only the constraint bounds ($load_e$) vary.
This property enables the \textbf{\emph{warm-start}} of the LPP solving by reusing the immediate states of the previous solution, significantly reducing optimization overhead.

\subsection{Routing Tokens to Replicas}
\label{ssec:routing}
Our next step is routing tokens to specific expert replicas, enforcing the calculated replica loads ($x_e^g$).
Specifically, we determine this \emph{token-to-replica} mapping using a sequential routing strategy:
First, we arrange tokens assigned to expert $e$ from all GPUs in sequence, along with an ordered list of expert $e$'s replicas.
Then, we iterate through these tokens, routing each to the first replica that has not yet reached its allocated load $x_e^g$.

To enhance efficiency, we can manipulate token ranges rather than individual tokens.
Let $input_e^g$ denote the input load of expert $e$ from GPU $g$, i.e., the number of tokens on GPU $g$ assigned to expert $e$ ($\sum_{g\in G_{\epname}}input_e^g=load_e$).
The pseudo code of token routing is shown in Algorithm~\ref{algo:route}, Lines 10-16.

\parab{Locality-aware routing.} 
Previous research has underscored that all-to-all communication introduces significant overhead to MoE systems~\cite{FasterMoE, Tutel, Lina}.
To reduce the all-to-all communication volume, we can leverage \emph{data locality} during token routing.
Specifically, when GPU $g$ holds a replica of expert $e$, we prioritize routing tokens from GPU $g$ to its local expert replica before considering remote replicas.
The pseudo code is shown in Algorithm~\ref{algo:route}, Lines 4-9.

\parab{Communication-aware scheduling.} Furthermore, we can consider communication overhead during scheduling.
Specifically, we can reformulate the optimization problem to minimize the overall maximum execution time, incorporating both computation and communication.
Moreover, we can model the heterogeneity between intra-node and inter-node communication.
Due to space limits, the modified optimization problem is detailed in Appendix~A.1.

\begin{algorithm}[t]
    \caption{Routing tokens to expert replicas.}
    \label{algo:route}
    \SetKwIF{If}{ElseIf}{Else}{if}{}{elif}{else}{}
    \KwIn{$\{input_e^g\}$, $\{x_e^g\}$}
    $\{remain\_input_e^g\}=\{input_e^g\}$\\
    $\{remain\_x_e^g\}=\{x_e^g\}$\\
    \For{$e\in E$}{
        // First, route local tokens to local replicas. \\
        \For{$g\in G_{EDP}^e$}{
            $y=\min(remain\_input_e^g, remain\_x_e^g)$\\
            route the next $y$ tokens of expert $e$ from GPU $g$ to the replica on GPU $g$\\
            $remain\_input_e^g\gets remain\_input_e^g-y$ \\
            $remain\_x_e^g\gets remain\_x_e^g-y$\\
        }
        // Then, route global tokens to global replicas. \\
        \For{$g\in G_{\epname}$}{
            \For{$g'\in G_{EDP}^e$}{
                $y=\min(remain\_input_e^g, remain\_x_e^{g'})$\\
                route the next $y$ tokens of expert $e$ from GPU $g$ to the replica on GPU $g'$\\
                $remain\_input_e^g\gets remain\_input_e^g-y$ \\
                $remain\_x_e^{g'}\gets remain\_x_e^{g'}-y$\\
            }
        }
    }
\end{algorithm}

\subsection{Distributed Scheduling across Devices}\label{ssec:where}
As described in \S\ref{ssec:lpp} and \S\ref{ssec:routing}, the scheduling algorithm requires global load information ($input_e^g$) across the entire \epname group to generate a token dispatching plan.
This raises the question of where to place the scheduler.

We consider two candidate locations for the scheduler: \emph{centralized} on one device or \emph{distributed} across all devices.
A centralized scheduler needs to gather load information from all devices, perform the scheduling, and scatter the results.
Alternatively, distributed schedulers require all devices to perform an all-gather operation to collect global load information and execute the scheduling algorithm independently.
\footnote{The distributed scheduling approach maintains consistency because \epname's scheduling algorithm is deterministic.
}

We choose the distributed approach for better scalability because it requires only one communication operation compared to two operations in the centralized approach.
Since the load information is small, the primary performance factor is latency rather than throughput, making fewer communication operations advantageous.

\subsection{Overlapping Scheduling to Hide Latency}\label{ssec:when}
Since we need to execute token scheduling in every micro-batch, minimizing the scheduling overhead is significant for system efficiency.
Therefore, we reduce the scheduling overhead by \textbf{\emph{overlapping}} scheduling with other operations.

As described in \S\ref{ssec:overview}, the scheduling executes immediately after the gate network and before the all-to-all communication.
We observe that existing distributed training frameworks usually perform some operations during this period.
For instance, Megatron-LM~\cite{Megatron-LM} executes a token permutation operation to replicate tokens by top-K times and sort tokens by expert indices before all-to-all dispatching.
Therefore, we can overlap the scheduling on CPUs with the permutation on GPUs.
For frameworks without suitable overlapping operations, we propose a \emph{pipelining} mechanism to hide scheduling latency, which is detailed in Appendix~A.2.


\section{Expert Placement}
\label{sec:placement}
Expert placement is fundamental to the load balancing performance of token scheduling in \epname.
\figurename~\ref{fig:flow-1} shows the relationship between expert placement and token scheduling in MicroMoE.
Specifically, according to LPP~\ref{eq:optimization}, the expert placement determines the EDP groups ($G_{EDP}^e$), which are the key components of the constraint matrix and thus significantly affect the optimization result.
Although our experiments demonstrate that some simple placement methods, such as random shuffling, usually provide acceptable results (\S\ref{ssec:vary_skew}), identifying optimal placement strategies is still crucial for maximizing system performance.
In the following sections, we first analyze ``\emph{What is an optimal expert placement?}'' in \S\ref{ssec:placement_analysis} and then discuss ``\emph{How to construct an optimal expert placement?}'' in \S\ref{ssec:symmetric} and \S\ref{ssec:asymmetric}.
Finally, we propose an \emph{adaptive replacement} mechanism in \S\ref{ssec:adaptive}.

\subsection{Analysis of Optimal Expert Placement}
\label{ssec:placement_analysis}

\begin{figure}[t]
    \centering
    \includegraphics[width=0.9\linewidth]{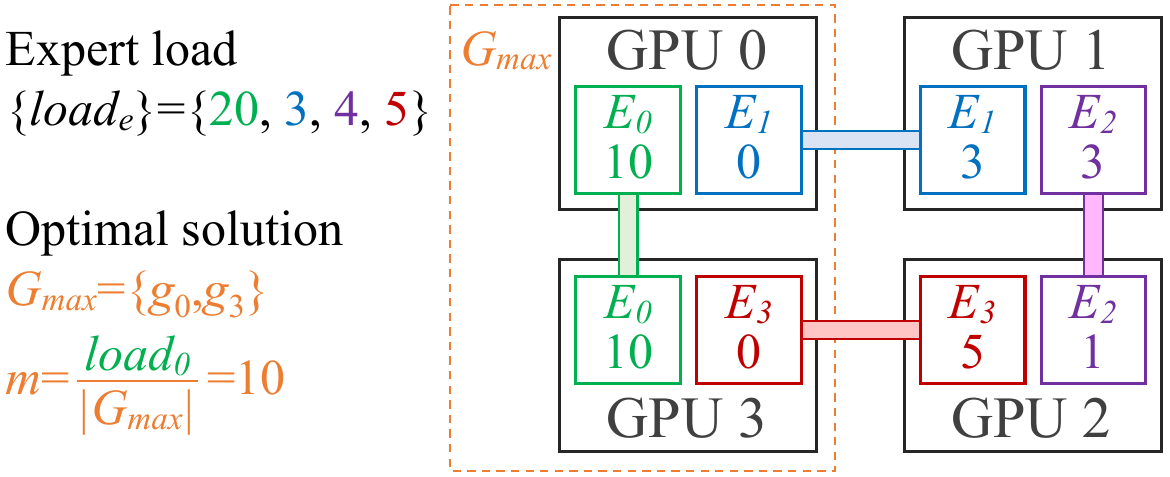}
    \caption{The graph abstraction of an example expert placement.
    Color bars indicate edges (experts) between vertices (GPUs).
    $G_{max}$ contains GPU 0, 3.
    Expert 0 is entirely in $G_{max}$. Experts 1,3 partially intersect with $G_{max}$ and cannot distribute any load within $G_{max}$.}
    \label{fig:graph}
\end{figure}


\parab{Analysis of the LPP solution.}
Since the optimal expert placement should minimize the objective value of LPP~\ref{eq:optimization}, we begin by analyzing the solution of this problem.
Let $m$ denote the optimal objective value of LPP~\ref{eq:optimization}, which represents the minimized maximum load across all GPUs.

\noindent\emph{Lemma.}
\label{lem:1} 
There exists an optimal solution to LPP~\ref{eq:optimization} that satisfies:
Let $G_{max}$ denote the set of GPUs with load $m$ in this solution.
An expert distributes no load within $G_{max}$ unless its EDP group is entirely in $G_{max}$.

\noindent\emph{Proof.} 
Among all optimal solutions, we choose the solution with the minimal $|G_{max}|$ (number of GPUs with load $m$).
Consider experts whose EDP groups partially intersect with $G_{max}$.
Suppose that the EDP group of expert $e$ consists of two GPUs $g_0\in G_{max}$ and $g_1\notin G_{max}$.
If expert $e$ did distribute any load on GPU $g_0$, it could move a tiny amount of load from $g_0$ to $g_1$.
Afterward, the loads of both $g_0$ and $g_1$ would be less than $m$, yielding a new optimal solution with smaller $|G_{max}|$. 
The new solution would contradict our choice of the solution with the minimal $|G_{max}|$.
Therefore, expert $e$ cannot distribute any load on GPU $g_0$.
$\hfill\qedsymbol$

\figurename~\ref{fig:graph} shows an example of the above lemma.
From this lemma, we can derive an equation by calculating the total loads within $G_{max}$ in two ways:
\begin{equation}
m\cdot |G_{max}|=\sum\limits_{e\in E:G_{EDP}^e\subseteq G_{max}}load_e\label{eq:m_1}
\end{equation}

From Equation~\ref{eq:m_1}, we can derive an equation to calculate $m$ by enumerating every possible $G_{max}$\footnote{We omit the derivation from Equation~\ref{eq:m_1} to Equation~\ref{eq:m_2}.}:
\begin{equation}
m=\max_{G_{max}\subseteq G_{\epname}}\left\{\frac 1 {|G_{max}|}\sum\limits_{e\in E:G_{EDP}^e\subseteq G_{max}}load_e\right\}\label{eq:m_2}
\end{equation}

Equation~\ref{eq:m_2} provides a mathematical approach to determine the optimal objective value of LPP~\ref{eq:optimization} without 
actually executing the LPP solving.

\parab{Graph abstraction of expert placement.}
Furthermore, we formulate expert placement with graph theory.
Let each GPU $g\in G_{\epname}$ represent a vertex, and each expert $e\in E$ represent a hyperedge connecting all GPUs in $G_{EDP}^e$.
Hence, an expert placement can be represented by an undirected hypergraph denoted by $\mathcal{G}(G_{\epname}, \{G_{EDP}^e:e\in E\})$. (When $d=2$, the hypergraph is a conventional graph. For simplicity, we omit the prefix ``hyper'' in the rest of this paper.)
\figurename~\ref{fig:graph} shows an example of the graph and the optimal solution corresponding to an expert placement.

Now, we can explain Equation~\ref{eq:m_2} from the perspective of graph theory:
We assign the weight of the edge (expert) $e$ as $load_e$.
We define the \emph{density} of a weighted graph as the sum of all edge weights divided by the number of vertices\footnote{This definition may differ from conventional graph density definitions in the literature.}.
Consider the subgraph in $\mathcal{G}$ induced by $G_{max}$.
The edges in the induced subgraph represent the experts whose EDP groups are entirely in $G_{max}$ ($\{e\in E:G_{EDP}^e\subseteq G_{max}\}$).
According to Equation~\ref{eq:m_2}, the optimal objective value $m$ is the maximum density across all induced subgraphs of graph $\mathcal{G}$.
Since the goal of expert placement is to minimize $m$, we can characterize the optimal expert placement as follows:

\parab{Property of the optimal expert placement.} 
\emph{The optimal expert placement is the graph whose maximum induced subgraph density is minimal.}

Having understood the property of the optimal expert placement, our next challenge is how to find such a placement.
We distinguish between two scenarios based on our knowledge of the expert load $load_e$ in \S\ref{ssec:symmetric} and \S\ref{ssec:asymmetric}.

\subsection{Symmetric Placement without Expert Loads}
\label{ssec:symmetric}
If we have no prior knowledge of the real expert load distribution, we can construct \emph{symmetric placements}, treating all experts equally.
Symmetric placements provide conservative and general load balancing capability in terms of unknown load distributions.
In such scenarios, we can assume that all expert loads follow an independent and identically distributed (i.i.d.) pattern.
Thereby, the problem becomes: Given the number of vertices and edges as well as the identical distribution of edge weights, how to construct a graph that minimizes the expectation of the maximum induced subgraph density?

We recognize the challenge of the above problem due to the vast space of possible graphs and distributions.
Nevertheless, we propose a near-optimal symmetric placement strategy for many practical configurations using \emph{Cayley graphs}~\cite{cayley}.
Our intuition is that the inherent symmetry of Cayley graphs ensures a balanced distribution of edges across vertices, preventing some induced subgraphs from having significantly larger density than others.
Since Cayley graphs involve complex group theory, we illustrate our construction method in Appendix B.




\subsection{Asymmetric Placement with Expert Loads}
\label{ssec:asymmetric}
If we know real expert load distributions in advance, we can construct \emph{asymmetric placements} tailored to them.
Unlike symmetric scenarios, we can vary both replica counts and replica locations across different experts for better load balancing, similar to previous works~\cite{SmartMoE, FlexMoE, Prophet}.

We adopt an empirical strategy to construct a near-optimal asymmetric expert placement in two steps:
(1) First, we determine the number of replicas for each expert 
with a \emph{greedy} algorithm: We maintain a heap of experts sorted by load-per-replica, and iteratively allocate remaining replicas to the expert with the maximum load-per-replica.
(2) Second, we determine the placement of expert replicas across GPUs with \emph{Monte Carlo sampling}: We randomly generate many placement graphs, and choose the one whose maximum induced subgraph density is minimal.

\begin{figure}[t]
    \centering
    \begin{subfigure}{\linewidth}
    \includegraphics[width=\linewidth, trim=0 170 0 0, clip]{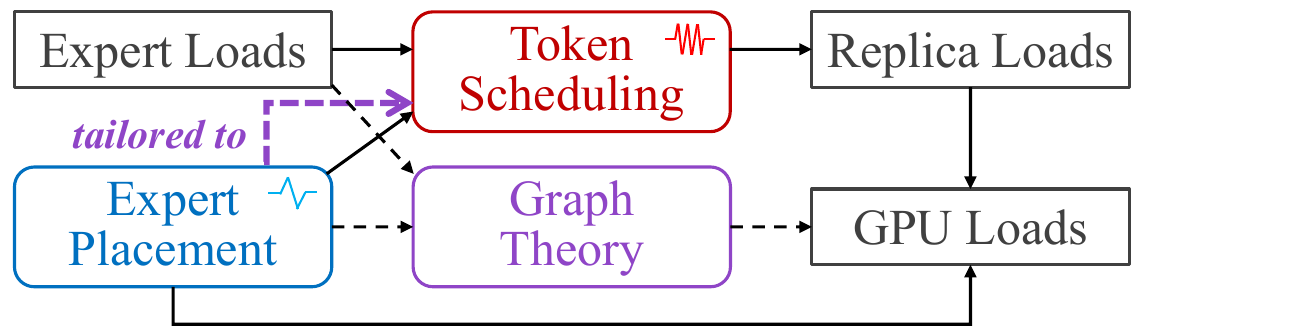}
    \caption{Data flow and analysis flow of MicroMoE.}
    \label{fig:flow-1}
    \end{subfigure}
    \begin{subfigure}{\linewidth}
    \includegraphics[width=\linewidth, trim=0 0 0 200, clip]{figs/flow.pdf}
    \caption{Data flow of expert scheduling systems.}
    \label{fig:flow-2}
    \end{subfigure}
    \caption{Differences between MicroMoE and expert scheduling systems.}
    \label{fig:flow}
\end{figure}

\subsection{Adaptive Replacement}
\label{ssec:adaptive}
Based on symmetric and asymmetric placements, we further propose an \emph{adaptive replacement} (AR) mechanism for \sysname to optimize performance under dynamic expert loads.

\parab{Relationship between token scheduling and adaptive replacement.}
The adaptive replacement mechanism complements the token scheduling in \S\ref{sec:schedule} by addressing different levels of load balancing.
Token scheduling performs \emph{transient, fine-grained} load balancing through per-micro-batch token arrangement, while adaptive replacement handles \emph{long-term, coarse-grained} load imbalances through periodic expert arrangement.

Specifically, for moderately imbalanced workloads, token scheduling sufficiently maintains complete balance with static, symmetric placements (as shown in \S\ref{ssec:vary_skew}).
For highly skewed workloads, \sysname adopts asymmetric placements to mitigate coarse-grained imbalances before using token scheduling for fine-grained optimization.
Since asymmetric placements require real-time expert loads, \sysname adopts adaptive replacement to monitor expert load distributions and adjust placements when significant distributional shifts are detected.

\parab{Implementation of adaptive replacement.} We implement the adaptive replacement mechanism in \sysname using the \emph{placement manager} (according to \figurename~\ref{fig:system}).
(1) During model initialization, the placement manager initializes the model states of all devices using the symmetric placement strategy, providing conservative and general load balancing capabilities.
(2) During training, the placement manager monitors expert load information within each micro-batch in the background.
(3) For every few iterations, the placement manager predicts future load distributions using historical data with time series analysis techniques, such as moving averages~\cite{Prediction}.
Then, it evaluates the performance of current placements on future distributions using Equation~\ref{eq:m_2}.
If the future performance drops below a specific threshold, the placement manager generates new optimal asymmetric placements and re-initializes global model states accordingly.

\parab{Difference between \sysname's adaptive replacement and expert scheduling solutions.}
The system implementation of \sysname's adaptive replacement is similar to existing expert scheduling solutions, such as FlexMoE~\cite{FlexMoE} and SmartMoE~\cite{SmartMoE}.
Nevertheless, their design goals and algorithms are fundamentally different:

\emph{Design goals}: In systems like FlexMoE, changing expert placement is the only means for load balancing, as shown in \figurename~\ref{fig:flow-2}.
However, in \sysname, the primary weapon is token scheduling, while adaptive replacement is a further optimization to token scheduling, as shown in \figurename~\ref{fig:flow-1}.
Even with static placement, \sysname can still achieve good load balancing performance at micro-batch granularity.

\begin{figure*}[t]
    \centering
    \includegraphics[width=0.85\linewidth, trim=10 30 10 10]{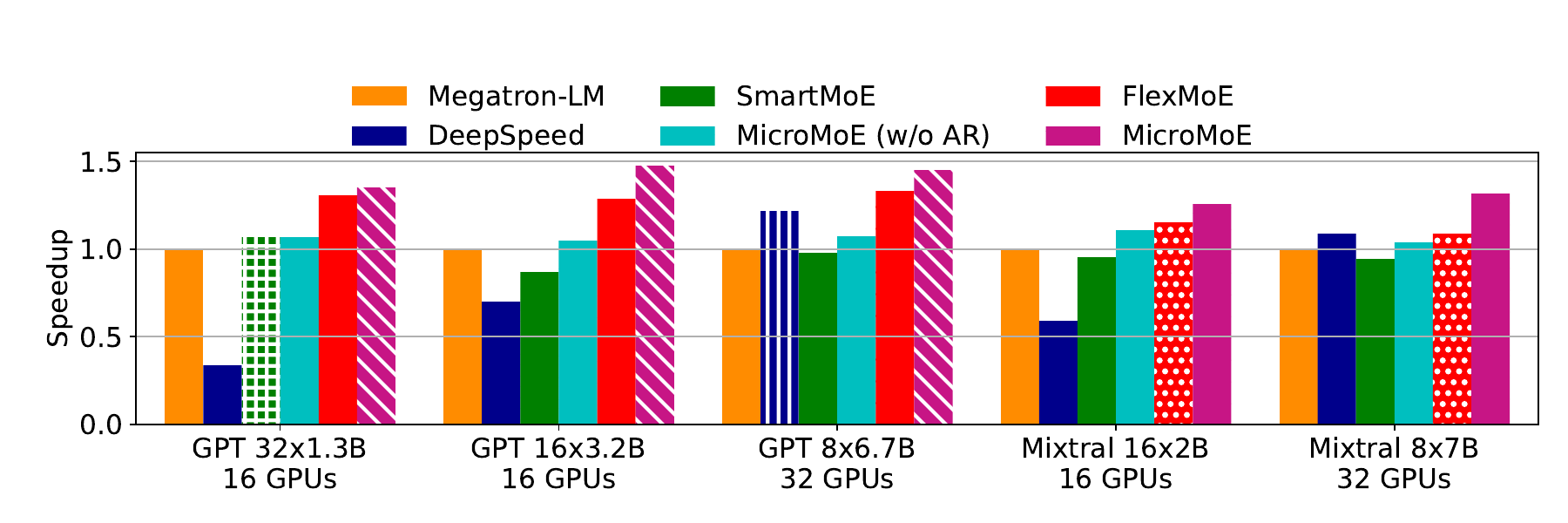}
    \caption{End-to-end speedup of different systems compared with Megatron-LM.}
    \label{fig:e2e}
\end{figure*}

\emph{Algorithms}: Different design goals yield distinct placement algorithms.
In existing expert scheduling systems, an expert's replicas typically have identical loads (i.e., replica\_load=expert\_load/replica\_count).
Therefore, they can leverage greedy or dynamic programming algorithms accordingly~\cite{SmartMoE, FlexMoE}.
In contrast, replica loads in \sysname are determined by linear programming.
Therefore, \sysname requires the graph theory in \S\ref{ssec:placement_analysis} to guide the placement strategy.

\section{Evaluation}
\label{sec:evaluation}
\subsection{Experimental Setup}

\parab{Testbed.}
Our testbed consists of 4 nodes, each equipped with 8 NVIDIA H100 80GB SXM GPUs connected via 900 GBps NVLink. 
Nodes are interconnected using two 400 Gbps Infiniband NICs per node.

\parab{Models.}
We use GPT~\cite{GPT-3} and Mixtral~\cite{Mixtral} models for evaluation.
We use GPT 32$\times$1.3B to represent an MoE model converted from a 1.3B dense GPT model with 32 experts.
We pretrain these models with the Wikipedia dataset~\cite{wiki}.
A detailed list of model hyperparameters is provided in Appendix~C. 

\parab{Baselines.}
We compare \sysname with four baseline systems.
\begin{itemize}
    \item Megatron-LM~\cite{Megatron-LM} is a popular distributed training framework for large language models (LLMs).
    It supports various parallelism strategies, as well as state-of-the-art optimizations~\cite{recomputation}.
    \item SmartMoE~\cite{SmartMoE} balances GPU loads by adjusting the expert placement within EP groups. 
    \item FlexMoE~\cite{FlexMoE} achieves load balancing by dynamically adjusting replica counts based on expert loads.
    FlexMoE places expert replicas across the entire DP group, similar to \sysname's asymmetric placement (\S\ref{ssec:asymmetric}).
    \item DeepSpeed~\cite{DeepSpeed} is a high-performance distributed framework for both LLM training and inference.
    We enable ZeRO-1~\cite{ZeRO} optimization in DeepSpeed (currently, ZeRO-2 does not support PP, and Tutel~\cite{Tutel} does not support top-K>1).
\end{itemize}

We compare two variants of \sysname: 
``\sysname (w/o AR)'' uses static, symmetric placement (\S\ref{ssec:symmetric}), while ``\sysname'' uses adaptive, asymmetric placement (\S\ref{ssec:asymmetric}-\ref{ssec:adaptive}).

\parab{Implementation.}
We implement \sysname upon Megatron-LM~\cite{Megatron-LM}.
\sysname provides a model wrapper similar to Pytorch's Distributed Data Parallel (DDP)~\cite{DDP}, enabling users to benefit from \epname's fine-grained load balancing capabilities within their training jobs.
We modify Megatron-LM, including its MoELayer and DDP with Python, and implement the token scheduling algorithm in \epname with C++.

For fair comparison, we also implement SmartMoE and FlexMoE in Megatron-LM, as SmartMoE's repository is outdated (last commit in 2023)~\cite{SmartMoE}, and FlexMoE is not open-sourced~\cite{FlexMoE}. 

\parab{Parallelization configurations.}
Due to the limited inter-node network bandwidth in our testbed, we only employ PP for inter-node parallelism.
Specifically, we set the PP degree to the number of nodes used, and the DP degree to 8.
We set the EP degree to 4, resulting in 2 EP groups per DP group.
We set the parameter $d$ in \epname to 2, resulting in a single \epname group per DP group.
We disable TP due to its high communication overhead.

\parab{Other configurations.}
We use a small auxiliary loss (listed in the appendix) to prevent extreme load imbalance from degrading model accuracy.
We enable the distributed optimizers in Megatron-LM, which resembles DeepSpeed's ZeRO-1~\cite{ZeRO}.
We disable the token dropping mechanism introduced by GShard~\cite{GShard}.
We use \texttt{BF16} precision.

\subsection{End-to-end Performance}
\label{ssec:e2e}
\figurename~\ref{fig:e2e} shows the end-to-end performance of all systems, varying models and number of GPUs.
DeepSpeed exhibits poor performance with 16 or 32 experts.
This is because DeepSpeed always adopts a padding mechanism, padding the load of each expert to the maximum expert load~\cite{GShard}.
This mechanism wastes significant time and memory when expert loads are highly imbalanced.
With as few as 8 experts, the inefficiency of padding becomes less significant, allowing DeepSpeed to outperform Megatron-LM due to its system-level optimizations.

Comparing SmartMoE and \sysname (w/o AR), where experts have uniform replica counts, \sysname (w/o AR) exhibits superior performance, thanks to \epname's fine-grained token scheduling.
While SmartMoE attempts to optimize performance by changing expert placement for load balancing among GPUs, it sometimes performs worse than vanilla Megatron-LM.
This is because SmartMoE optimizes expert placement based on long-term expert load distributions.
However, expert loads are highly dynamic during training, and SmartMoE's long-term optimal placement may be sub-optimal for individual micro-batches.

Comparing FlexMoE and \sysname (with AR), where experts have varied replica counts, \sysname exhibits superior performance, thanks to its fine-grained, per-micro-batch token scheduling.

In conclusion, \textbf{compared with Megatron-LM, \sysname improves the end-to-end throughput by up to 47.6\%, with an average improvement of 36.9\%.} 
The average performance improvement of \sysname surpasses FlexMoE, the second-best system, by 13.9\%.
Note that this is already the upper bound of performance improvement attainable through load balancing, since MicroMoE already achieves complete balance (detailed in \S~\ref{ssec:vary_skew}).

\begin{figure}[t]
    \centering
    \includegraphics[width=\linewidth, trim=10 30 10 10]{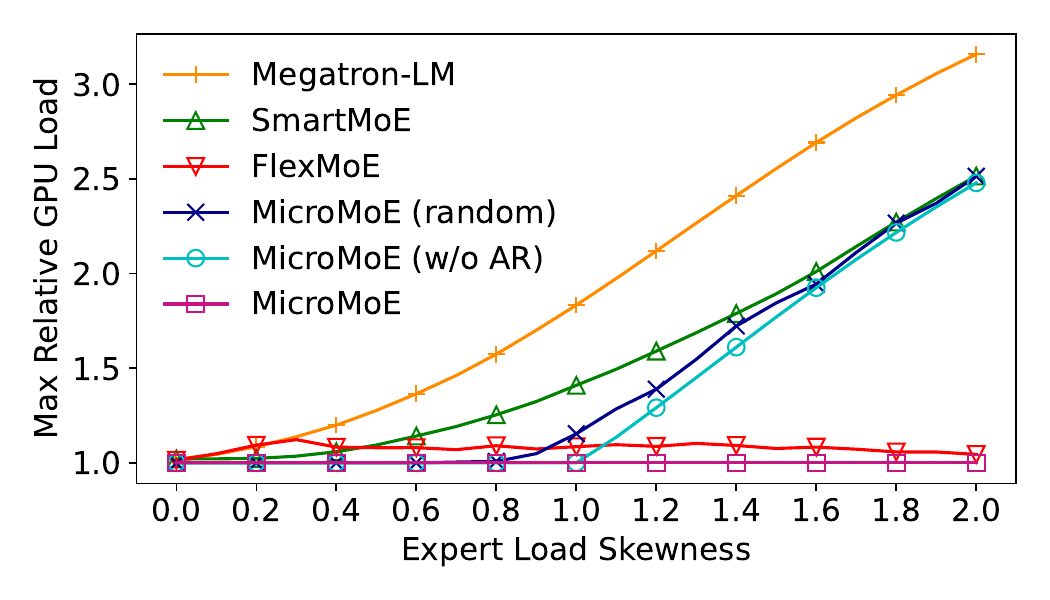}
    \caption{Max GPU load normalized by average GPU load, varying skewness of expert loads. DP\_degree=8, num\_experts=32.}
    \label{fig:vary_skew}
\vspace{-10pt}
\end{figure}

\subsection{Load Balancing Capability}
\label{ssec:vary_skew}
We evaluate the load balancing capabilities of SmartMoE, FlexMoE, and \sysname with skewed expert loads.
We generate expert loads following a Zipfian distribution with skewness $s$, where the probability of a token being assigned to the $i$-th most loaded expert is proportional to $i^{-s}$.

\figurename~\ref{fig:vary_skew} shows the load balancing performance of different systems across varied skewness $s$.
SmartMoE's maximum GPU load increases as load skewness increases.
While FlexMoE maintains relatively balanced GPU loads by adjusting expert replica counts, it falls short of achieving optimal load balance due to its lack of fine-grained dynamicity.
\sysname (random) represents \sysname with pure random placement, which performs slightly worse than \sysname (w/o AR) with symmetric placement.
\sysname (w/o AR) achieves perfect load balance when $s<1$ thanks to \epname's fine-grained token scheduling. 
Nonetheless, its performance degrades when $s>1$ as uniform replica counts are insufficient for severe imbalances.
\sysname with asymmetric placements can always achieve complete load balance, due to the combination of both coarse-grained expert replacement and fine-grained token scheduling.
Overall, \textbf{\sysname exhibits the best load balancing capability among all systems and consistently achieves complete load balance.}


\subsection{Execution Time Breakdown}
\label{ssec:breakdown}
\begin{figure}
    \centering
    \includegraphics[width=\linewidth, trim=10 30 10 10]{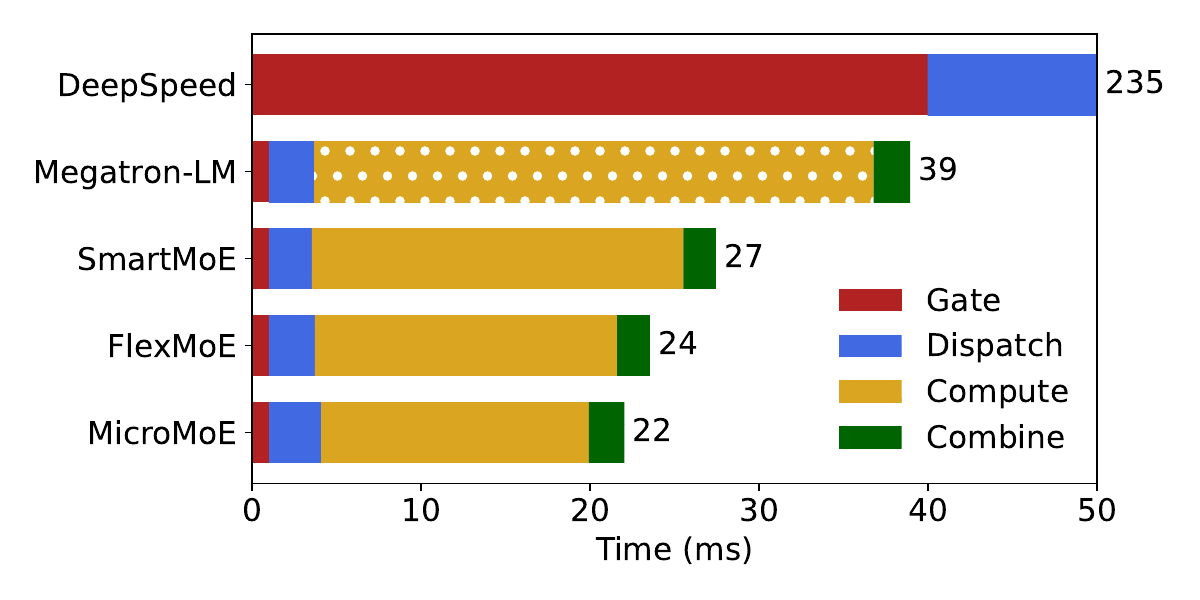}
    \caption{Execution time breakdown of an MoE layer. DP\_degree=8, num\_experts=32, micro\_batch\_size=8, sequence\_length=2048, topK=2, hidden\_size=4096, skewness $s$=1.}
    \label{fig:breakdown}
\vspace{-10pt}
\end{figure}

\figurename~\ref{fig:breakdown} shows the execution time breakdown of an MoE layer across different systems.
We omit DeepSpeed due to its poor performance.
For all remaining systems, the primary bottleneck is expert computation time.
\sysname achieves the shortest computation time by maintaining perfect load balance (with either symmetric or asymmetric placement, as shown in \figurename~\ref{fig:vary_skew}).

Specifically, the dispatch time consists of two primary components: (1) Preparation time, which includes the all-gather operation of expert load information and the scheduling of \epname.
While \sysname introduces additional overhead in dispatch time due to token scheduling operations, we effectively minimize this impact through overlapping with other operations in Megatron-LM.
(2) All-to-all communication time.
Each all-to-all communication in dispatch and combine requires approximately 1.3 ms in Megatron-LM.


\begin{figure*}
\begin{minipage}[t]{0.35\linewidth}
    \centering
    \includegraphics[width=\linewidth, trim=10 30 10 10]{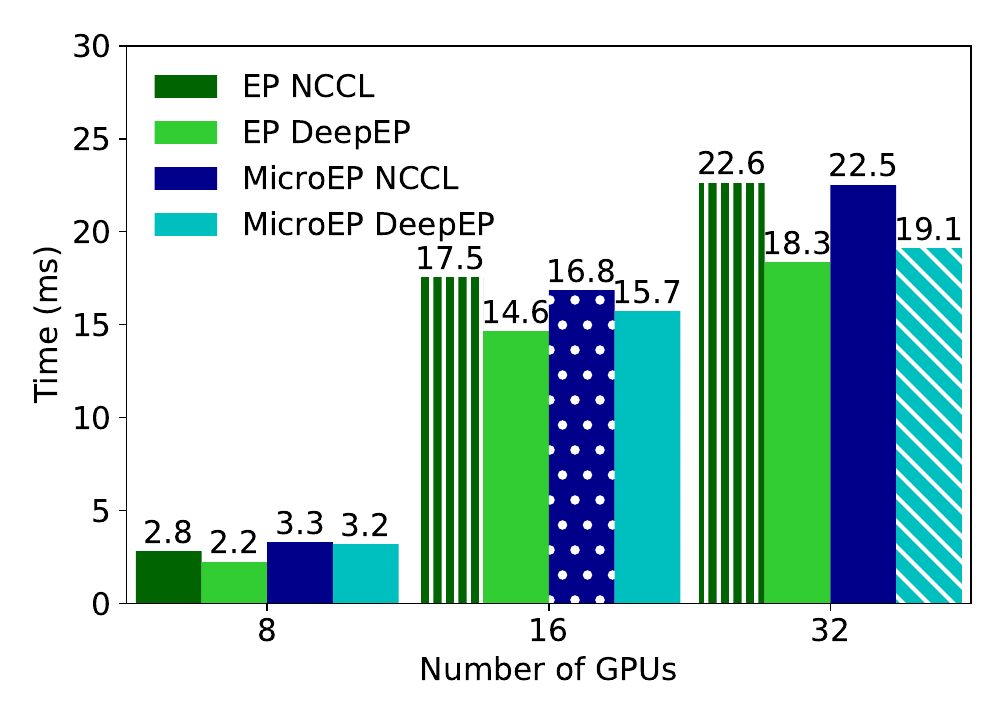}
    \caption{Dispatch time of \epname and EP with DeepEP and NCCL, varying number of GPUs.}
    \label{fig:deepep}
\end{minipage}
\begin{minipage}[t]{0.31\linewidth}
    \centering
    \includegraphics[width=\linewidth, trim=10 30 10 10]{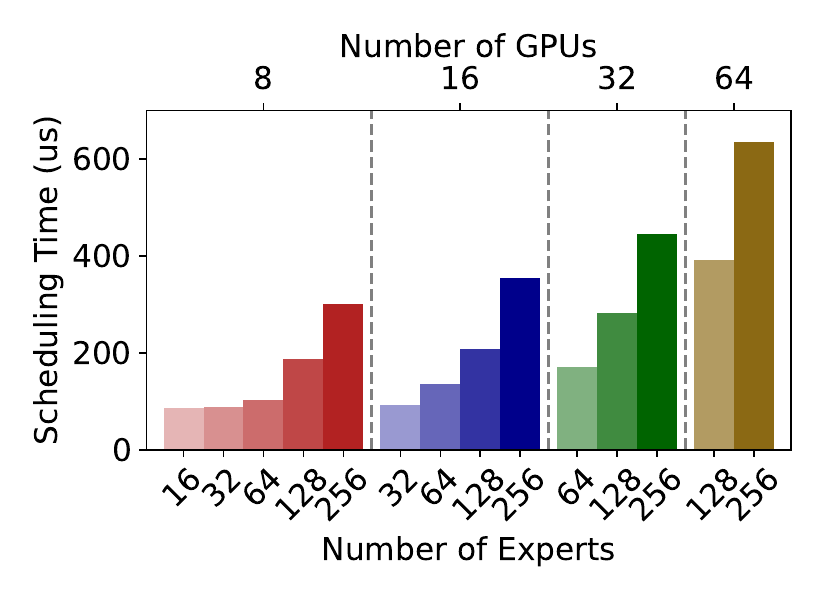}
    \caption{Scheduling time for \epname, varying number of experts and GPUs.}
    \label{fig:solve_time}
\end{minipage}
\begin{minipage}[t]{0.33\linewidth}
    \centering
    \includegraphics[width=\linewidth, trim=10 30 10 10]{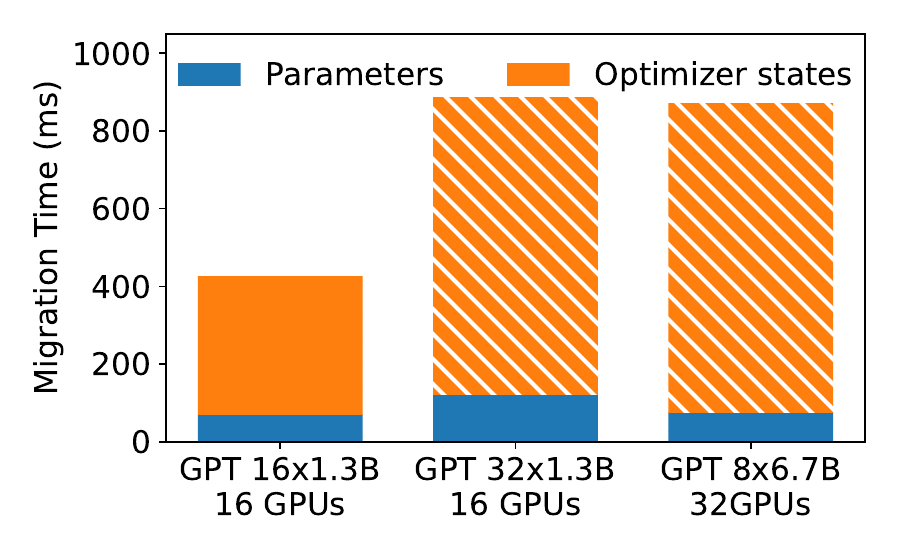}
    \caption{Migration time for adaptive replacement of \sysname.}
    \label{fig:migrate}
\end{minipage}
\vspace{-10pt}
\end{figure*}
\subsection{Inter-node Communication with DeepEP}
\label{ssec:deepep}
We evaluate the dispatch time of \epname and vanilla EP for inter-node communication.
We additionally integrate \epname with DeepEP~\cite{DeepSeek-V3}, a high-performance all-to-all communication backend.
Currently, Megatron-LM~\cite{Megatron-LM} supports both NCCL~\cite{NCCL} (by default) and DeepEP for all-to-all communication.

\parab{Experimental considerations.} Due to testbed limitations, two important experimental considerations should be noted:
(1) Our testbed consists of 8 GPUs but only 2 NICs per node, resulting in limited inter-node network bandwidth.
Therefore, we avoid employing EP or \epname across multiple nodes in \S\ref{sec:evaluation}.
However, since this section focuses on evaluating the performance of different communication backends, we expand the communication group to multiple nodes.
Consequently, the all-to-all time for inter-node communication is significantly higher than the intra-node communication.
(2) \S\ref{sec:evaluation} focuses on system performance, so we compare \epname using 8 GPUs per group with EP using 4 GPUs per group ($d=2$).
However, this section focuses on communication performance, so we compare \epname and EP using the same group size.

\parab{Overhead of inter-node communication.} Notably, since \epname expands the all-to-all group size by a factor of $d$, \epname may convert some inter-node communication into inter-node, leading to extra overhead.
However, this overhead is minimal in two typical scenarios: 
(1) When $d\times \text{EP\_degree}\le$ \# GPUs per node, the all-to-all communication in \epname remains entirely intra-node.
(2) When EP\_degree is super large (e.g., 64 in DeepSeek-v3~\cite{DeepSeek-V3}), nearly all communication is inherently inter-node.
Consequently, \epname incurs negligible overheads in both \emph{single-node} and \emph{massive-node} scenarios.
Furthermore, our communication-aware scheduling mechanism can jointly optimize the time of both communication and computation, as shown in Appendix C.2.

\parab{Results.} \figurename~\ref{fig:deepep} shows the dispatch time comparison between \epname and EP using both DeepEP and NCCL, varying number of GPUs.
We use the same setting as in \S\ref{ssec:breakdown}, except for the all-to-all group size.
DeepEP exhibits better performance than NCCL due to its superior implementation.
When using NCCL, \epname requires less time than EP, thanks to the locality-aware routing in \S\ref{ssec:routing}.
However, when using DeepEP, \epname requires more time than EP due to data format incompatibilities between DeepEP and Megatron-LM.
Consequently, Megatron-LM needs to pre-process the data for DeepEP, while \epname incurs a higher pre-processing overhead than EP.
We believe that \sysname will yield lower communication overheads on other practical testbeds (e.g., with one NIC per GPU).

\subsection{Overhead Analysis}
\label{ssec:overhead}

\parab{Scheduling Overhead.}
We evaluate the scheduling overhead of \epname, including the LPP solving time and token routing time.
Our evaluation reveals that the LPP solving time is the dominant factor, which scales with the number of experts and GPUs.
As shown in \figurename~\ref{fig:solve_time}, the scheduling overhead remains remarkably low, with a minimum time of approximately 100 us.
Even with 64 GPUs and 256 experts, the scheduling time remains below 1~ms.
This minimal overhead per micro-batch enables \epname to maintain high training throughput while providing load balancing benefits.
Additionally, we evaluate the performance of pipelining to hide the scheduling latency, which results are shown in Appendix~C.3.

\parab{Replacement overhead.}
The adaptive replacement strategy in \sysname necessitates model re-initialization to transition to new configurations.
Although replacement is beneficial for load balancing, model re-initialization causes temporary suspension of training.
Our evaluation highlights two key components in the replacement overhead: the migration time of expert parameters and their optimizer states.
As shown in \figurename~\ref{fig:migrate}, the total migration time typically spans hundreds of milliseconds across different model configurations.

The above results emphasize the importance of carefully selecting the expert replacement frequency to optimize the trade-off between per-micro-batch training efficiency and overall replacement overhead.
In practice, we recommend tuning the replacement interval to 50 iterations during the beginning phase of training, which adds less than 1\% overhead to the entire system.
One may increase this interval to several hundred iterations or even make no replacement when workloads become less volatile, as shown in \figurename~\ref{fig:dist}.

\begin{figure}
    \centering
    \includegraphics[width=\linewidth, trim=10 30 10 10]{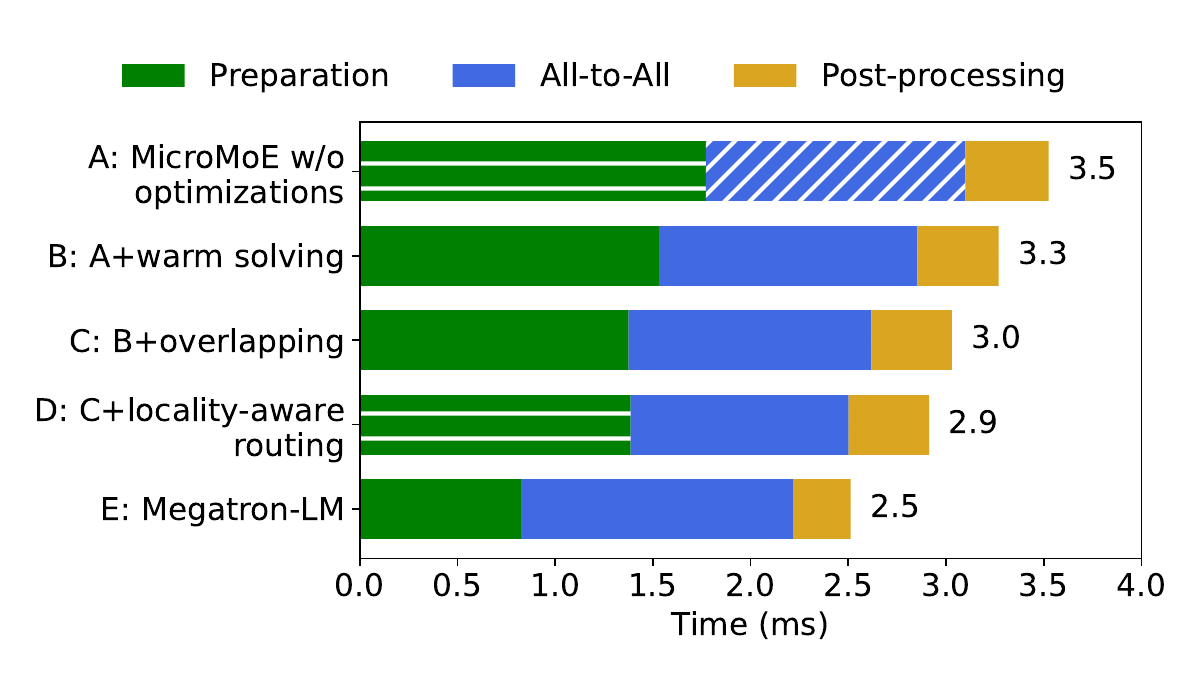}
    \caption{Ablation study of dispatch time, with the same setting as \figurename~\ref{fig:breakdown}.}
    \label{fig:ablation}
\end{figure}
\subsection{Ablation Study}
\label{ssec:ablation}
%
We conducted an ablation study to evaluate three optimizations in \sysname: (1) warm solving in \S\ref{ssec:lpp}, (2) locality-aware routing in \S\ref{ssec:routing}, and (3) overlapping in \S\ref{ssec:when}.
All these optimizations aim to reduce the dispatch time. 
As shown in \figurename~\ref{fig:ablation},
both warm solving and overlapping reduce scheduling time, while locality-aware routing reduces all-to-all communication time.
With the combination of optimizations, \sysname introduces only 0.4 ms additional dispatch time compared to vanilla Megatron-LM.
This modest overhead is substantially outweighed by the reduced computation time, ensuring the overall system efficiency of \sysname.

\section{Related Work}

\parab{MoE training systems.}
Switch Transformer~\cite{Switch_Transformer} pioneers expert parallelism for transformer-based models. 
FairSeq~\cite{FairSeq} and FastMoE~\cite{FastMoE} provide PyTorch~\cite{PyTorch} plugins to optimize MoE training.
DeepSpeed-MoE~\cite{DeepSpeed-MoE} combines EP with various optimizations, such as ZeRO~\cite{ZeRO} for large-scale MoE training.
Tutel~\cite{Tutel} dynamically adapts parallelism strategies and pipeline degrees to handle dynamic workloads of MoE.

\parab{MoE load balancing.}
Many studies propose algorithmic solutions to address the load imbalance issue of MoE.
Shazeer et al.~\cite{shazeer2017outrageouslylargeneuralnetworks} introduce a load-balancing loss, which is further refined by subsequent works~\cite{Switch_Transformer, LocMoE, auxiliary-loss-free, attention_router, TA-MoE, DSelect-k}.
GShard~\cite{GShard} introduces the concept of expert capacity to limit the number of tokens assigned to each expert, which is adopted by later works~\cite{Switch_Transformer, LocMoE, Tutel}.
Some studies propose alternative gating algorithms in place of the original top-K gating strategy~\cite{BASE, expert_choice, Hash, LSH-MoE}.
As a systematic approach, \sysname is compatible with all these algorithmic solutions.

Many studies offer systematic solutions for load balancing in MoE.
FasterMoE~\cite{FasterMoE} models the time consumption of MoE and enables broadcasting the most popular expert to all devices.
MoESys~\cite{MoESys} dynamically adjusts the number of training nodes to ensure load balance in multi-task scenarios. 
SmartMoE~\cite{SmartMoE} focuses on optimizing expert placement.
FlexMoE~\cite{FlexMoE} adapts the number of replicas of each expert based on their popularity.
Several other works~\cite{Hecate, SwiftMoE, Pro-Prophet, Prophet} embrace similar approaches to FlexMoE, and further reduce the adjustment overhead by integrating expert migration with ZeRO's communication operations~\cite{ZeRO}.
While these solutions typically balance loads at a per-iteration, expert-level granularity, \sysname is capable of achieving optimal load balance at a per-micro-batch granularity.

LPLB~\cite{LPLB} is a recent work that leverages linear-programming-based token scheduling for MoE load balancing, similar to MicroMoE. 
However, it lacks comprehensive analysis and experimental evaluation. \footnote{By the time we wrote this paper, LPLB was still in the early research stage and had no evaluation results. As far as we know, LPLB is the only relevant work that schedules tokens for load balancing.}
In contrast, this paper theoretically analyzes the relationship between expert placement and token scheduling, proposes multiple placement strategies, and incorporates key optimizations such as locality-aware routing.
We believe that our work provides clearer insights and practical guidance for designing next-generation MoE systems.

Janus~\cite{Janus} proposes a data-centric EP paradigm that avoids the inherent load imbalance issue in conventional expert-centric paradigms.
However, Janus requires migrating all expert parameters within every micro-batch, which incurs substantial communication overhead, especially when dealing with large batch numbers and model scales.

\section{Discussion}

\parab{Scalability.}
Scaling \sysname to large clusters and increased expert counts introduces new challenges to the system.
When scaling up, \sysname benefits from an expanded scheduling space, which enhances the load balancing capability of \epname.
However, this larger scheduling space increases computational overhead in scheduling.
To compromise between load balancing capability and system efficiency, we can organize GPUs and experts into groups and perform scheduling at the group level, similar to prior works~\cite{SmartMoE, GShard, DeepSeek-V3}.

\parab{FSDP.}
We currently implement \sysname based on Megatron-LM's DDP~\cite{Megatron-LM}.
Megatron-LM also supports Fully Sharded Data Parallel (FSDP)~\cite{FSDP}, which resembles DeepSpeed's ZeRO-3~\cite{ZeRO}, sharding model parameters and gradients for memory saving.
We plan to integrate \sysname with FSDP in future work.

\section{Conclusion}
We propose \epname, a novel expert parallelism strategy to achieve fine-grained load balancing in MoE.
\epname dynamically balances GPU loads within every micro-batch through token scheduling.
We primarily make two optimizations in \epname:
First, we formulate the token scheduling process as a linear programming problem, which can be solved efficiently.
Second, we theoretically analyze the relationship between expert placement and load balancing capacity and develop two placement strategies for different training scenarios.
Finally, we propose \sysname, an efficient MoE training system based on \epname.
Our experimental evaluation demonstrates that \sysname achieves significant performance improvements, with up to 47.6\% end-to-end speedup compared to Megatron-LM, and almost consistently maintains optimal load balance across GPUs.

\newpage

\bibliographystyle{plain}
\bibliography{reference}

\appendix
\section{Optimizations to \epname}
\label{app:optimization}

\subsection{Communication-Aware Scheduling}
\label{ssec:communication}
Many studies demonstrate that the all-to-all communication also takes a significant amount of time in MoE layers~\cite{Tutel, FasterMoE, Lina}.
However, the optimization problem~\ref{eq:optimization} only considers the computation time.
Therefore, we can further consider the communication time during scheduling.



In many existing frameworks~\cite{Megatron-LM, DeepSpeed}, the time of an MoE layer is mainly determined by the communication time + the computation time (unless using some communication-computation overlapping techniques, such as Comet~\cite{Comet} and DualPipe~\cite{DeepSeek-V3}).
Therefore, we introduce an additional objective to minimize the maximum communication volume on a GPU.
According to Algorithm~\ref{algo:route}, Line 6, the local data volume of GPU $g$ can be formulated as $local_g=\sum\limits_{e\in E: g\in G_{EDP}^e}\min(x_e^g, input_e^g)$.
For each GPU, we consider its all-to-all communication volume as the greater value between its send volume and receive volume.
The send/receive volume of GPU $g$ can be calculated from $x_e^g, input_e^g$, and $local_g$.
Putting them together, we derive the new optimization problem:
\begin{equation}
\begin{aligned}
\text{minimize} \quad & comp+\alpha \cdot comm, \\
\text{subject to} \quad & comp = \max\limits_{g\in G_{\epname}}\left\{\sum\limits_{e\in E: g\in G_{EDP}^e}x_e^g\right\}, \\
& comm = \max\limits_{g\in G_{\epname}}\left\{\max(send_g, recv_g)\right\}, \\
& send_g = \left(\sum\limits_{e\in E: g\in G_{EDP}^e}input_e^g\right)-local_g, \\
& recv_g = \left(\sum\limits_{e\in E: g\in G_{EDP}^e}x_e^g\right)-local_g, \\
& local_g=\sum\limits_{e\in E: g\in G_{EDP}^e}\min\left(x_e^g, input_e^g\right), \\
& \sum\limits_{g\in G_{EDP}^e}x_e^g=load_e, \quad \forall e\in E, \\
& x_e^g \ge 0.
    \label{eq:communication-1}
\end{aligned}
\end{equation}

$\alpha$ is a constant parameter reflecting the weight of the communication time in the optimization problem.
In practice, we can set $\alpha$ to the ratio of expert computation throughput to all-to-all communication throughput.
The optimization problem~\ref{eq:communication-1} is still an LPP.
We omit the derivation of converting problem~\ref{eq:communication-1} to a standard LPP format. 

\parab{Topology-aware scheduling.}
We can further consider the impact of network topology on communication.
In a typical network topology, intra-node communication (e.g., through NVLink) is many times faster than inter-node communication (e.g., through Infiniband).
Therefore, we can consider this difference during scheduling for better communication efficiency.


In the LPP~\ref{eq:communication-1}, we can split the communication time into intra-node and inter-node.
Intra-node communication has a lower weight ($\alpha_1$) than inter-node communication ($\alpha_2$).
Additionally, we modify the routing mechanism as follows:
First, route local tokens to local replicas within the same GPUs.
Second, route tokens to replicas on other GPUs in the same nodes.
Third, route tokens to other global GPUs.

\parab{Overhead.}
While communication-aware scheduling minimizes communication volume, it increases computational complexity due to additional parameters and constraints in the LPP (comparing LPP~\ref{eq:communication-1} and \ref{eq:optimization}). 
Therefore, we should carefully consider when to enable the communication-aware scheduling, performing a trade-off between the LPP solving time and the communication time.

\subsection{Pipelining \epname}
\label{ssec:pipeline}
\begin{figure}[ht]
    \centering
    \begin{subfigure}{\linewidth}
    \includegraphics[width=\linewidth, trim=0 200 0 10, clip]{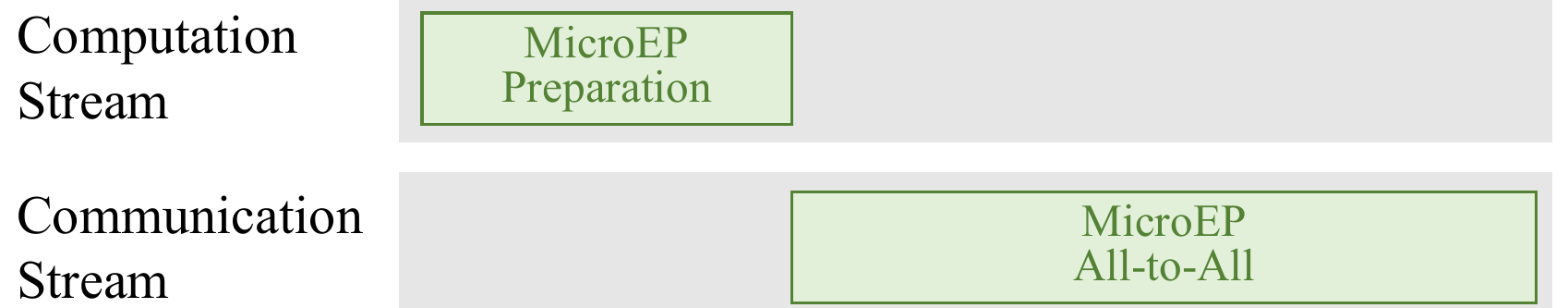}
    \caption{\epname without pipelining.}
    \label{fig:pipeline-1}
    \end{subfigure}
    \begin{subfigure}{\linewidth}
    \includegraphics[width=\linewidth, trim=0 10 0 200, clip]{figs/pipeline.pdf}
    \caption{\epname with pipelining.}
    \label{fig:pipeline-2}
    \end{subfigure}
    \caption{Pipelining \epname.}
    \label{fig:pipeline}
\end{figure}
Pipelining offers an additional approach to hide the scheduling overhead of \epname, different from the overlapping technique in \S\ref{ssec:when}.
Pipelining is particularly useful in scenarios where it is impossible to fully hide the scheduling time through overlapping.
For example, some high-performance frameworks, such as DeepEP~\cite{DeepSeek-V3}, have no intermediate operations to overlap between the gate network and all-to-all communication; when dealing with large-scale deployments with numerous GPUs and experts, the scheduling can incur substantial overhead (as illustrated in \S\ref{ssec:overhead}).
With pipelining, we can overlap the scheduling of some tokens with the all-to-all communication of other tokens.
This approach provides more flexibility in latency hiding.

We find many design choices to perform pipelining for \epname and illustrate one of them as follows.
We observe that \epname can achieve perfect load balance in most circumstances (as shown in \S\ref{ssec:vary_skew}).
Our intuition is that even if the expert load becomes more imbalanced, \epname can still achieve perfect balance.
Inspired by this observation, we can split the tokens into two parts:
We apply EP to the former part\footnote{Since we have already changed expert placement, this EP is somehow different from typical EP and more like FlexMoE.}, and apply \epname to the latter part.
Moreover, the optimization problem~\ref{eq:optimization} should consider the computational workloads in both EP and \epname.
In this way, we can overlap the scheduling of the latter part with the all-to-all communication of the former part, as shown in \figurename~\ref{fig:pipeline}.

\parab{Overhead.}
Pipelining \epname introduces some additional system overhead.
For example, splitting one all-to-all communication operation into two increases synchronization time and GPU kernel launching time.
Therefore, we recommend using \epname with pipelining in scenarios with substantial scheduling time but minimal system overhead.

\section{Expert Placement using Cayley Graphs}
\label{app:cayley}
\subsection{Cayley Graphs}
In many practical applications, we can construct near-optimal symmetric expert placements in \epname using Cayley graphs.
The inherent symmetry of Cayley graphs makes them well-suited for constructing optimal expert placements.

A Cayley graph is constructed from a \emph{group} $A$ and its \emph{generating set} $S$.
Each element in the group $a\in A$ is assigned as a vertex.
For every $a\in A$ and $s\in S$, there is an edge from the vertex $a$ to the vertex $as$.

We assume the \epname parameter $d=2$, so the hypergraph is a conventional graph.
We observe that in practical applications, the quantities of GPUs and experts are usually powers of two.
Let the number of GPUs be $2^p$, and the number of experts per GPU be $2^{q}$.
Consequently, the number of vertices is $2^p$, the degree of each vertex is $2^q$, and the number of edges is $2^{p+q-1}$.
In practice, we heuristicly construct many Cayley graphs for different $(p, q)$s.

\begin{figure}
    \centering
    \begin{subfigure}{0.6\linewidth}
    \includegraphics[width=\linewidth]{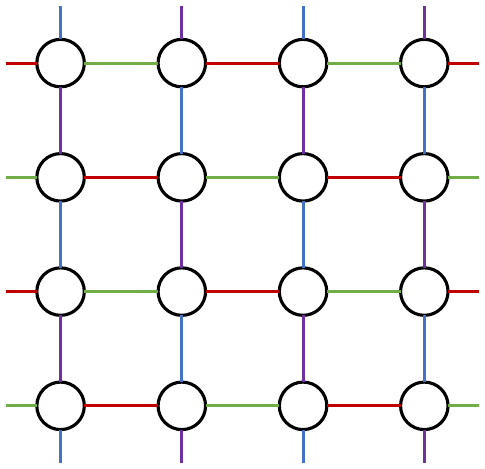}
    \caption{16 vertices, 32 edges.}
    \label{fig:cayley-1}
    \end{subfigure}
    \begin{subfigure}{0.7\linewidth}
    \includegraphics[width=\linewidth]{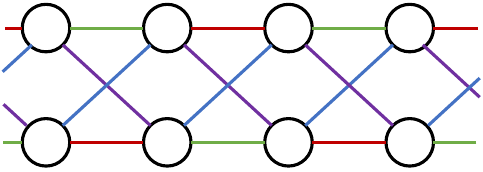}
    \caption{8 vertices, 16 edges.}
    \label{fig:cayley-2}
    \end{subfigure}
    \caption{Examples of Cayley graphs.}
    \label{fig:cayley}
\end{figure}

\subsection{Examples of Cayley Graphs}
\label{ssec:example_cayley}
We illustrate some example constructions as follows.

\parab{Example 1:} 8 vertices, 8 edges.

We have $p=3, q=1$.
The group is $(\mathbb{Z}_8, +)$, and the generating set is $\{1, -1\}$.
The constructed graph is a cycle.

\parab{Example 2:} 16 vertices, 32 edges.

We have $p=4, q=2$. 
The group is $(\mathbb{Z}_4\times \mathbb{Z}_4, +)$, and the generating set is $\{(0,1), (0,-1), (1,0), (-1,0)\}$.
This graph is a $4\times 4$ toroidal grid graph, as shown in \figurename~\ref{fig:cayley-1}.


\parab{Example 3:} 8 vertices, 16 edges.

We have $p=3, q=2$. 
The group is $(\mathbb{Z}_2\times \mathbb{Z}_4, +)$, and the generating set is $\{(0,1), (0,-1), (1,1), (1,-1)\}$.
The constructed graph is shown in \figurename~\ref{fig:cayley-2}, which is isomorphic to the complete bipartite graph $K_{4, 4}$.

This construction satisfies a good property:
$\forall i\in 1,...,8$, the maximum edge counts among all induced subgraphs with exactly $i$ vertices is minimal.

\parab{Example 4:} 8 vertices, 32 edges.

Note that a complete graph with 8 vertices has $C_8^2=28$ edges.
Since complete graphs are certainly optimal, we can first generate a complete graph and then add the remaining $32-28=4$ edges.
For the remaining 4 edges, we can simply create an edge between every vertex pair (0, 1), (2, 3), (4, 5), (6, 7) without using Cayley theory.

This method is generalizable to scenarios with more edges: We can first generate multiple complete graphs and then allocate the remaining edges.
Note that the number of edges is a power of 2, the number of vertices is $2^p$, and the number of edges in a complete graph is $\frac {2^p(2^p-1)} 2$. 
Consequently, the number of remaining edges must still be a power of 2, ranging from $2^{p-1}$ to $2^{2p-2}$.

\subsection{Synchronization Consistency}
Different EDP groups across experts can lead to a consistency issue during parameter and gradient synchronization.
Specifically, the synchronization for different experts occurs in different EDP groups, which may incur deadlocks.
To prevent deadlocks, we add a consistency restriction in expert placement: All replicas of an expert must have identical local expert indices.
For example, in \figurename~\ref{fig:overview-3}, the replicas of expert 2 are the first local replicas of both GPU 1 and GPU 2; the colors of edges in \figurename~\ref{fig:cayley} also indicate the local expert indices.
Since DDP executes parameter synchronization following the order of local parameters (and gradient synchronization following the reverse order)~\cite{DDP}, deadlocks are effectively avoided.

\section{Supplementary Experiments}
\label{app:experiment}
\subsection{Detailed Experimental Settings}
\label{ssec:detail_setting}
This section provides detailed configurations for our experiments in \S\ref{sec:evaluation}.
Table~\ref{tab:model} lists the detailed hyperparameters for models used in \S\ref{ssec:e2e}.

\emph{Activation recomputation} is a technique to reduce memory footprint by avoiding recording activations in the forward pass and recomputing them in the backward pass~\cite{recomputation}.
Furthermore, \emph{selective activation recomputation} enables recomputing a subset of model modules to perform a fine-grained trade-off between computation efficiency and memory~\cite{Megatron-SP}. 
We enable selective recomputation in Megatron-LM to recompute only the MoE FFN, avoiding the out-of-memory (OOM) issue while maintaining relatively high throughput. 
Since DeepSpeed currently does not support selective recomputation, we recompute the whole layer in DeepSpeed.
Furthermore, we find that we can adjust the granularity of selective recomputation at runtime.
When the expert loads are highly imbalanced, we can recompute the whole MoE layer for better memory efficiency, avoiding OOM.
Otherwise, we can recompute only the MoE FFN for better computation efficiency.
For fair comparison, we do not adjust the recomputation granularity during evaluation.

\begin{table*}
  \caption{List of Model Hyperparameters}
  \label{tab:model}
  \begin{tabular}{rrrrrr}
    \toprule
    Model &
    GPT 32$\times$1.3B & GPT 16$\times$3.2B & GPT 8$\times$6.7B & Mixtral 16$\times$2B & Mixtral 8$\times$7B \\
    \hline
    \# layers & 24 & 16 & 32 & 32 & 32 \\
    \# attention heads & 16 & 32 & 32 & 32 & 32 \\
    hidden size & 2048 & 4096 & 4096 & 2048 & 4096 \\
    FFN hidden size & 8192 & 16384 & 16384 & 8192 & 14336 \\
    sequence length & 2048 & 2048 & 2048 & 4096 & 4096 \\
    \# experts & 32 & 16 & 8 & 16 & 8 \\
    top-K & 2 & 2 & 2 & 2 & 2 \\
    micro batch size & 4 & 2 & 2 & 2 & 1 \\
    global batch size & 512 & 512 & 512 & 256 & 256 \\
    learning rate & 1e-5 & 2e-6 & 1e-6 & 1e-5 & 1e-6 \\
    load-balancing loss coeff. & 1e-4 & 1e-4 & 1e-4 & 1e-4 & 5e-4 \\
    \# GPUs & 16 & 16 & 32 & 16 & 32 \\
    PP degree & 2 & 2 & 4 & 2 & 4 \\
    EP degree & 4 & 4 & 4 & 4 & 4 \\
  \bottomrule
\end{tabular}
\end{table*}

\subsection{Evaluation of DeepEP}
\label{ssec:deepep}
\begin{figure}[t]
    \centering
    \includegraphics[width=\linewidth, trim=10 30 10 10]{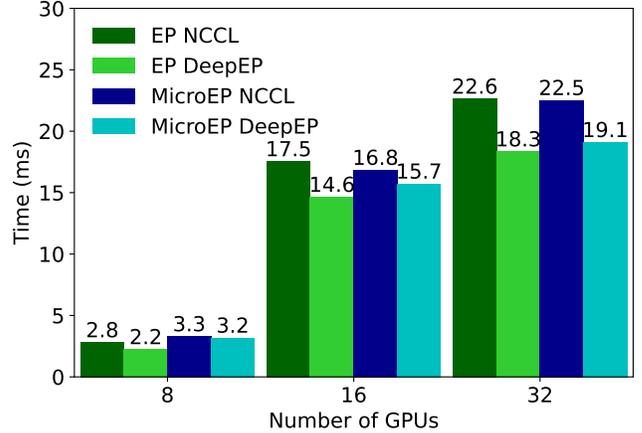}
    \caption{Dispatch time of \epname and EP with DeepEP and NCCL, varying number of GPUs.}
    \label{fig:deepep}
\end{figure}
We evaluate the dispatch time of \epname and vanilla EP with DeepEP~\cite{DeepSeek-V3}, a high-performance all-to-all communication backend.
Megatron-LM~\cite{Megatron-LM} currently supports both NCCL~\cite{NCCL} (by default) and DeepEP for all-to-all communication.
We additionally implement \epname with DeepEP.

Before analyzing the experimental results, two important experimental considerations should be noted:
(1) Our testbed consists of 8 GPUs and only 2 NICs per node, resulting in limited inter-node network bandwidth.
Therefore, we avoid employing EP or \epname across multiple nodes in \S\ref{sec:evaluation}.
However, in this section, we focus on the performance of different communication backends, so we expand the communication group into multiple nodes.
Consequently, the all-to-all time for inter-node communication is significantly longer than the intra-node communication.
(2) In \S\ref{sec:evaluation}, we focus on the system performance, so we compare \epname using 8 GPUs per group with EP using 4 GPUs per group ($d=2$).
However, in this section, we focus on the communication performance, so we compare \epname and EP using the same group size.

\figurename~\ref{fig:deepep} shows the dispatch time comparison between \epname and EP using both DeepEP and NCCL, varying number of GPUs.
We use the same setting as in \S\ref{ssec:breakdown}, except for the all-to-all group size.
DeepEP exhibits better performance than NCCL due to its high-performance all-to-all implementation.
When using NCCL, \epname requires less time than EP thanks to the locality-aware routing in \S\ref{ssec:routing}.
However, when using DeepEP, \epname requires more time than EP due to the data format incompatibilities between DeepEP and Megatron-LM.
Consequently, Megatron-LM needs to pre-process the data for DeepEP, while \epname incurs a higher pre-processing overhead than EP.

\subsection{Evaluation of Communication-Aware Scheduling}
\label{ssec:eval_comm}
\begin{figure}[t]
    \centering
    \includegraphics[width=\linewidth, trim=10 30 10 10]{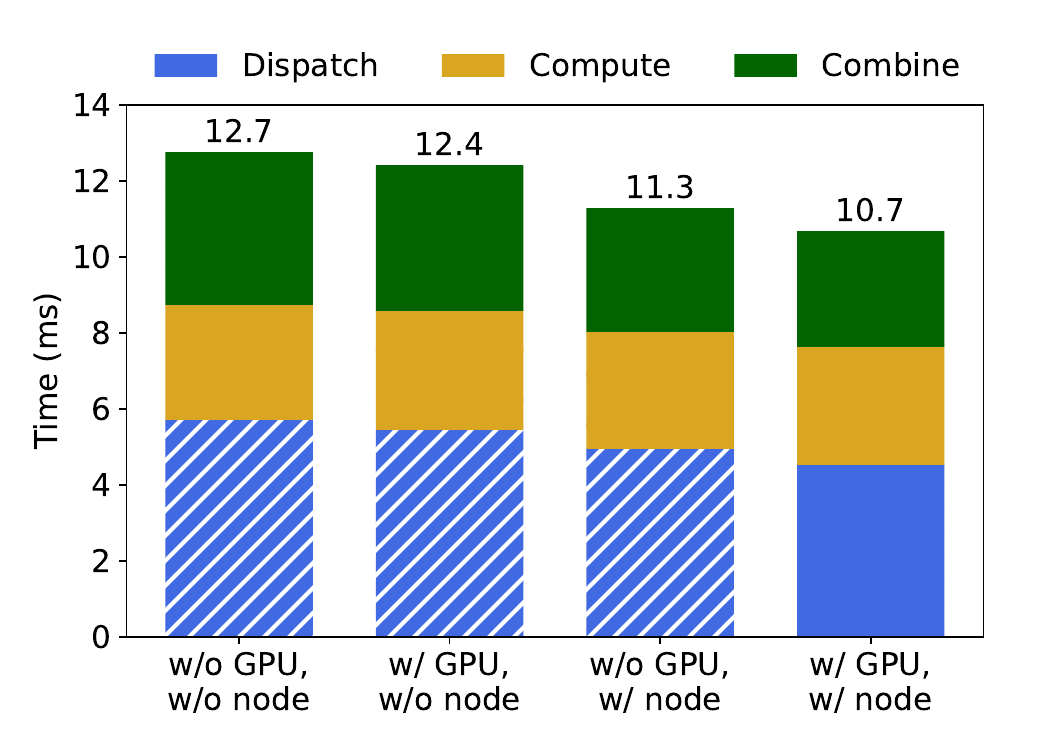}
    \caption{Execution time breakdown of an MoE layer with \epname, varying the levels of communication-aware scheduling.}
    \label{fig:communication}
\end{figure}
We evaluate the performance of the communication-aware scheduling in Appendix~\ref{ssec:communication}.
The communication-aware scheduling considers two levels of locality in token dispatching: GPU-level (intra-node) locality and node-level (inter-node) locality.
We set $\alpha_1=0.1, \alpha_2=1.0$ as the weights of intra-node and inter-node communication in Problem~\ref{eq:communication-1}.
We use DeepEP as the communication backend due to its superior performance and reduced system overhead compared to NCCL.
We believe that the dispatch time of DeepEP provides a more accurate reflection of the communication volume.
For other parameters, we use 16 GPUs, 32 experts, hidden\_size=2048, sequence\_length=4, micro\_batch\_size=4.
We use randomly generated tokens as input.

We compare the execution time of an MoE layer while enabling/disabling the GPU-level/node-level locality in the communication-aware scheduling.
As shown in 
\figurename~\ref{fig:communication}, the overall execution time decreases as we consider more levels of locality during scheduling.

\subsection{Evaluation of \epname with Pipelining}
\label{ssec:eval_pipe}
\begin{figure}[t]
    \centering
    \includegraphics[width=\linewidth, trim=10 30 10 10]{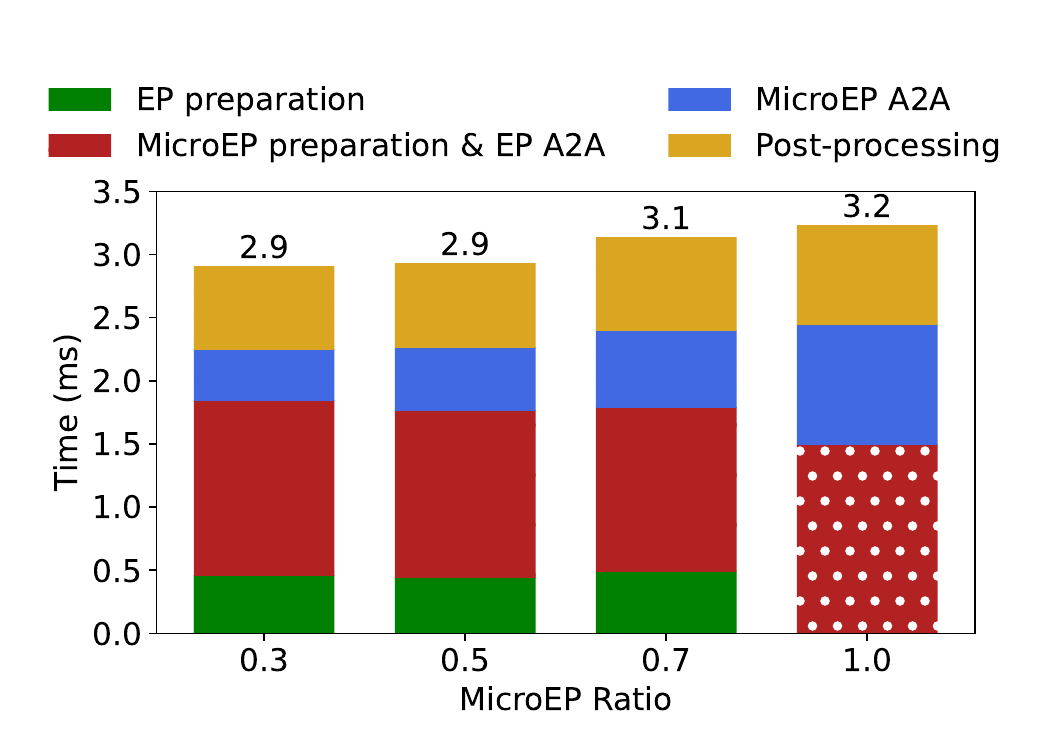}
    \caption{Dispatch time breakdown with pipelining, vary the ratios of data in \epname (1.0 indicates no pipelining).}
    \label{fig:pipe_ratio}
\end{figure}
We evaluate the performance of \epname with pipelining in Appendix~\ref{ssec:pipeline}.
We enable the communication-aware scheduling and DeepEP.
We use 8 GPUs and 128 experts. 
Other parameters are the same as Appendix~\ref{ssec:eval_comm}.

We compare the dispatch time with varying ratios of data in \epname.
\figurename~\ref{fig:pipe_ratio} demonstrates that pipelining can reduce the dispatch time by overlapping \epname preparation with EP all-to-all communication.
However, the dispatch time increases as the \epname ratio increases.
This is because the EP all-to-all time decreases and becomes insufficient to fully hide the \epname scheduling time.

\end{document}